\documentclass[english,twocolumn,amssymb,amsmath,superscriptaddress]{revtex4}
\usepackage[latin9]{inputenc}
\usepackage{graphicx,color}
\usepackage{floatrow}
\usepackage{subfig}
\usepackage{hyperref}
\usepackage{babel}
\usepackage{booktabs}
\usepackage{hyperref}
\usepackage[normalem]{ulem}
\usepackage{caption}
\newcommand{\be}{\begin{equation}}
\newcommand{\ee}{\end{equation}}

\definecolor{mygreen}{rgb}{0,0.5,0}
\definecolor{myblue}{rgb}{0,0,0.75}
\definecolor{mymagenta}{cmyk}{0,1,0,0.12}

\begin{document}

\title{Generalized transfer matrix states from artificial neural networks}

\author{Lorenzo Pastori}
\affiliation{Institute of Theoretical Physics, Technische Universit\"at Dresden, 01062 Dresden, Germany}
\affiliation{Department of Physics, University of Gothenburg, SE 412 96 Gothenburg, Sweden}
\author{Raphael Kaubruegger}
\affiliation{Department of Physics, University of Gothenburg, SE 412 96 Gothenburg, Sweden}
\affiliation{Center for Quantum Physics, Faculty of Mathematics, Computer Science and Physics, University of Innsbruck,  A-6020 Innsbruck, Austria}
\affiliation{Institute for Quantum Optics and Quantum Information of the Austrian Academy of Sciences, A-6020 Innsbruck, Austria}
\author{Jan Carl Budich}
\affiliation{Institute of Theoretical Physics, Technische Universit\"at Dresden, 01062 Dresden, Germany}	
\date{\today}

\begin{abstract}  
Identifying variational wave functions that efficiently parametrize the physically relevant states in the exponentially large Hilbert space is one of the key tasks towards solving the quantum many-body problem. Powerful tools in this context such as tensor network states have recently been complemented by states derived from artificial neural networks (ANNs). Here, we propose and investigate a new family of quantum states, coined generalized transfer matrix states (GTMS), which bridges between the two mentioned approaches in the framework of deep ANNs. In particular, we show by means of a constructive embedding that the class of GTMS contains generic matrix product states while at the same time being capable of capturing more long-ranged quantum correlations that go beyond the area-law entanglement properties of tensor networks. While the state amplitude of generic deep ANNs cannot be exactly evaluated, that of a GTMS network can be analytically computed using transfer matrix methods. With numerical simulations, we demonstrate how the GTMS network learns random matrix product states in a supervised learning scheme, and how augmenting the network by long-ranged couplings leads to the onset of volume-law entanglement scaling. By means of an explicit example using variational Monte Carlo, we also show that GTMS can parametrize critical quantum many-body ground states to a good accuracy. Our findings suggest that GTMS are a promising candidate for the study of critical and dynamical quantum many-body systems.   
\end{abstract}

\date{\today}

\maketitle

\section{INTRODUCTION}
\label{sectio:intro}
The quantum many-body problem is one of the outstanding challenges in physics. Besides providing deep theoretical insights, its solution may enable revolutionary technological applications including room-temperature superconductivity and new nano-technology enabled by the understanding of complex macro-molecules. The key issue in this context is the exponential complexity of generic quantum many-body states with the number of constituents. A widely applicable and successful approach towards taming this exploding complexity is to devise families of variational states that efficiently parametrize the physical scenario under investigation. 

\floatsetup[figure]{style=plain,subcapbesideposition=top}
\begin{figure} [h!]
	\sidesubfloat[]{\includegraphics[width=0.57\columnwidth]{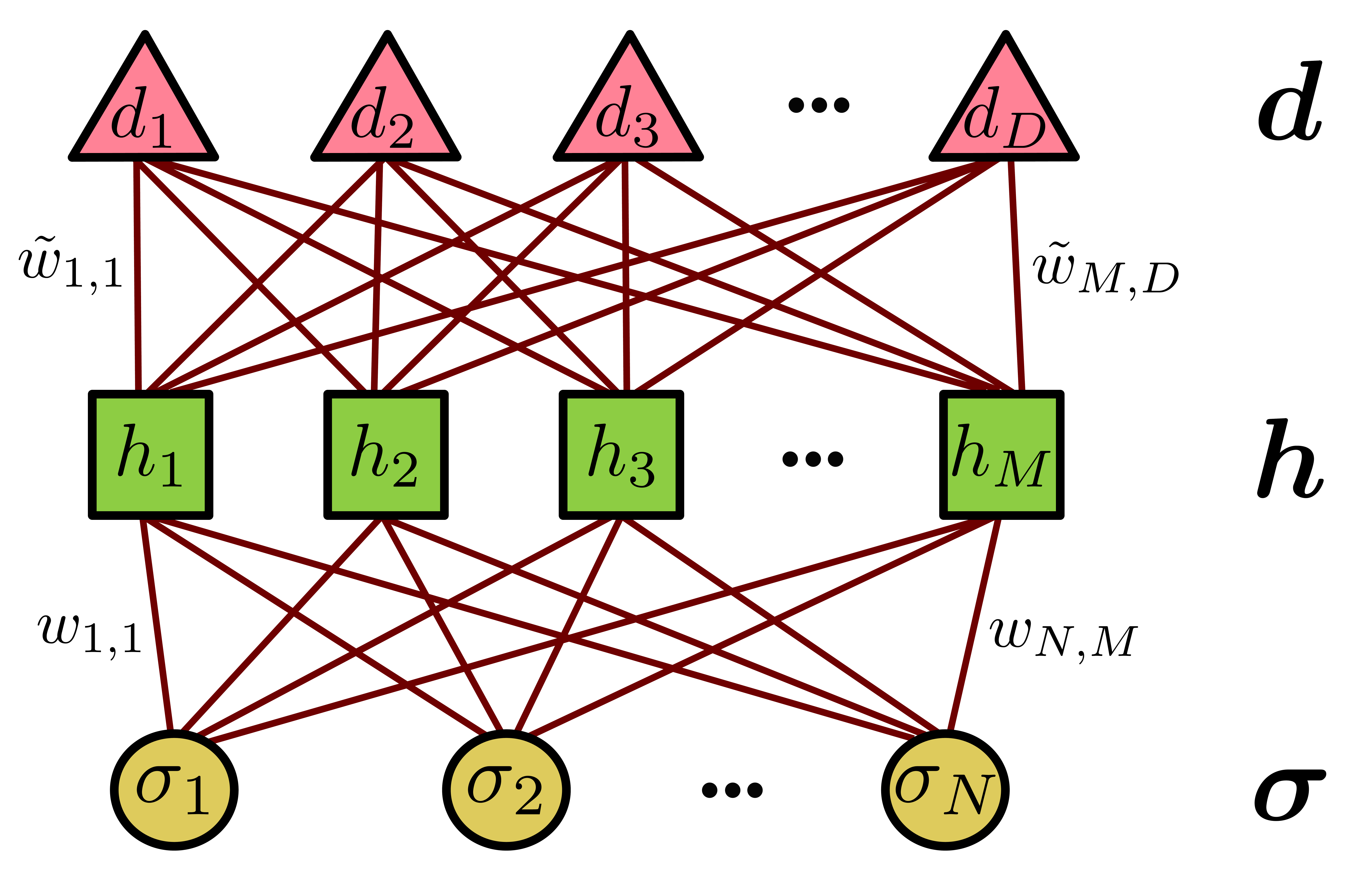}\label{fig:DBM_full}}\\
	\sidesubfloat[]{\includegraphics[width=0.98\columnwidth]{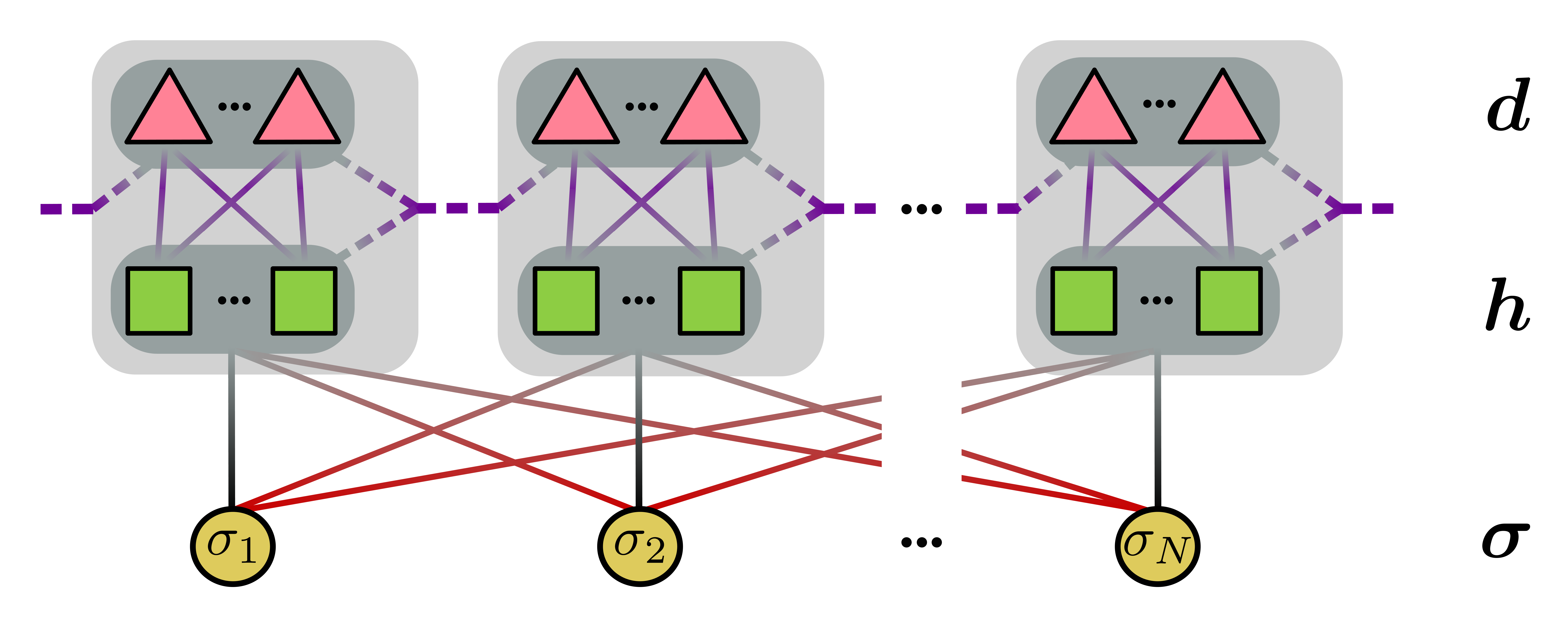}\label{fig:generalGTMS}}
	\caption{(color online) (a): Three-layer DBM architecture. Yellow circles denote the physical sites $\boldsymbol{\sigma}=\{\sigma_1,...,\sigma_N\}$, the green boxes the hidden units $\boldsymbol{h}=\{h_1,...,h_M\}$, and the pink triangles the deep units $\boldsymbol{d}=\{d_1,...,d_D\}$. Complex weights $w_{i,j}$ and $\tilde{w}_{j,k}$ are associated to the links between physical/hidden and hidden/deep layers, respectively. Not drawn in the picture are the bias terms. (b) Schematic representation of a GTMS network. Hidden and deep units are grouped in blocks (grey-shaded boxes) corresponding to transfer matrices. Keeping only black and purple links yields a MPS, while the red links make the transfer matrices dependent on all the physical spins. By cutting the purple links in the GTMS network, one retrieves a RBM.}
	\label{fig:DBM_GTMS}
\end{figure}

A paradigmatic example along these lines is provided by tensor network states such as matrix product states (MPS) \cite{WhiteDMRG,McCullochiDMRG,SchollwoeckMPS_DMRG,Ostlund_MPS_DMRG,Cirac_MPS_DMRG} and their higher-dimensional generalizations \cite{CiracPEPS,VidalTTN,NoackTTN,MarcelloSimoneTTN,Vidal_MERA,CiracReview_TNS}. The tensor network ansatz generally describes quantum correlations that are reflected in the area-law entanglement of the wave function, thus successfully capturing ground states of gapped local Hamiltonians \cite{EisertPlenio_AreaLaw,Hastings_AreaLaw,Vedral_entanglement_rev}. Tensor network states have also become an important tool for the study of critical systems \cite{Vidal_Kitaev_entangCritical,Schollwoeck_DMRG_Luttinger,Laflorencie_Entang1Dcrit} and dynamical properties \cite{Vidal_TimeEvo,White_TimeEvo,ZaletelExactMPS} even though it is clear that these situations exhibit quantum correlations beyond area-law entanglement. The price to pay for encompassing such scenarios is to increase the size of the tensors, i.e. the number of variational parameters, with system size and (exponentially) with time, respectively. Hence, the stronger growth of entanglement limits the amenable system sizes and periods of time evolution.

Complementing existing variational approaches \cite{WhiteDMRG,McCullochiDMRG,SchollwoeckMPS_DMRG,Ostlund_MPS_DMRG,Cirac_MPS_DMRG,CiracPEPS,VidalTTN,NoackTTN,MarcelloSimoneTTN,Vidal_MERA,CiracReview_TNS,McMillan_HeVMC,Ceperley_HeVMC,SorellaSR,CasulaSorellaSR,Jastrow,Gutzwiller}, quantum states derived from artificial neural networks (ANN) \cite{CarleoTroyer} have recently been introduced and studied. There, the physical degrees of freedom are coupled to a set of auxiliary units [see Fig.~\ref{fig:DBM_full}], and the wave function is obtained summing --- or tracing --- over all configurations of the auxiliary degrees of freedom forming the auxiliary layer(s), thus retaining the couplings as variational parameters. Analytically tracing over the auxiliary layer of one of the simplest ANN architectures known as restricted Boltzmann machine (RBM) \cite{Bengio_RBMrepresentability} already leads to quantum states \cite{CarleoTroyer} that exhibit volume-law entanglement \cite{DengVolumelaw}, thus offering an alternative variational wave function for those situations in which short range quantum correlations captured by conventional tensor network methods may not be sufficient \cite{DengVolumelaw,RBMStabilizer,NomuraHubbard,SaitoHubbard,our_article}. Using the more complex ANN class of deep Boltzmann machines (DBM) \cite{HintonDBM,GaoDBM,Bengio_RBMrepresentability} [see Fig.~\ref{fig:DBM_full}], it has recently been proven that the imaginary time evolution towards the ground state of a generic many-body Hamiltonian can be exactly represented at polynomial network complexity \cite{Carleo_exactImag}. However, this does not imply an efficient solution of a given many-body problem, since computing the sum over the auxiliary layers of a DBM in general is a exponentially hard problem. Hence, the explicit form of the wave function is in general not accessible even if it can be efficiently represented graphically in the DBM framework.

The purpose of this work is to develop a hybrid approach bridging between tensor network and ANN states. To this end, we introduce and study a class of DBM networks, which we coin generalized transfer matrix state (GTMS) networks [see Fig.~\ref{fig:generalGTMS}], where the auxiliary layers are exactly traceable, meaning that the wave function can be exactly analytically computed. There the wave function is analytically evaluated using transfer matrix methods. The resulting GTMS are capable of arbitrarily interpolating between MPS and RBM states thus combining key physical properties of these two powerful variational methods. As a limiting case, we obtain conventional MPS (RBM states) from the GTMS architecture by cutting the red (purple) couplings in Fig.~\ref{fig:generalGTMS}. To demonstrate how arbitrary MPS are efficiently parameterized in the proposed framework, we show that GTMS networks can indeed \emph{learn} random MPS by optimizing the coupling parameters in a supervised learning scheme. Furthermore, we argue how the GTMS generalizes and augments the class of MPS by making the tensors nonlocally dependent on the physical degrees of freedom. This more complex structure allows the GTMS to capture correlations beyond area-law entanglement. Our analysis is supported by numerical studies on the scaling of the second R\'enyi entanglement entropy, showing that with the addition of nonlocal neural couplings in the network [red links in Fig.~\ref{fig:generalGTMS}] the GTMS indeed acquires volume-law entanglement. This increased representational power makes the GTMS a promising candidate for the study of critical and time-dependent systems, which we numerically demonstrate in the case of critical systems by parametrizing the ground-state of the XXZ chain at the Kosterlitz-Thouless transition point with a variational Monte Carlo (VMC) calculation.

Several previous works have investigated the relationship between ANN sates and tensor network states \cite{HuangPiP,GlasserMunich,Clark,ChenANN_TN}, establishing a general correspondence between certain Boltzmann machine (among which short-range RBM) architectures and MPS representations of quantum states. Going beyond these previous insights, our present construction provides a constructive and efficient embedding of MPS and RBM states into the general framework of DBM networks. This allows us to continuously interpolate between MPS and RBM states and generalize both approaches in a physically motivated fashion.

The remainder of the paper is organized as follows. In the next section, we review the concept of a deep Boltzmann machine state. In Sec. \ref{sectio:GTMS}, we introduce the GTMS, discussing the DBM architectures which allows for exact traceability of the auxiliary layers by means of a transfer matrix approach, and showing how it can interpolate between MPS and RBM. In Sec. \ref{sectio:MPSfromGTMS}, we explicitly show that MPS are a special case of GTMS, and we construct the paradigmatic AKLT state from a GTMS network as an example. We also numerically show that the GTMS architecture can be trained to learn generic random MPS. In Sec. \ref{sectio:entanglementGTMS}, we provide a numerical analysis of the second R\'enyi entanglement entropy of the GTMS, and in Sec. \ref{sectio:XXZ}, we present our numerical result for the VMC optimization of GTMS for learning the ground-state of a critical XXZ chain. We finally present a concluding discussion in Sec. \ref{sectio:Conclusion}.

\begin{figure*}  [htp]
	\includegraphics[width=0.75\columnwidth]{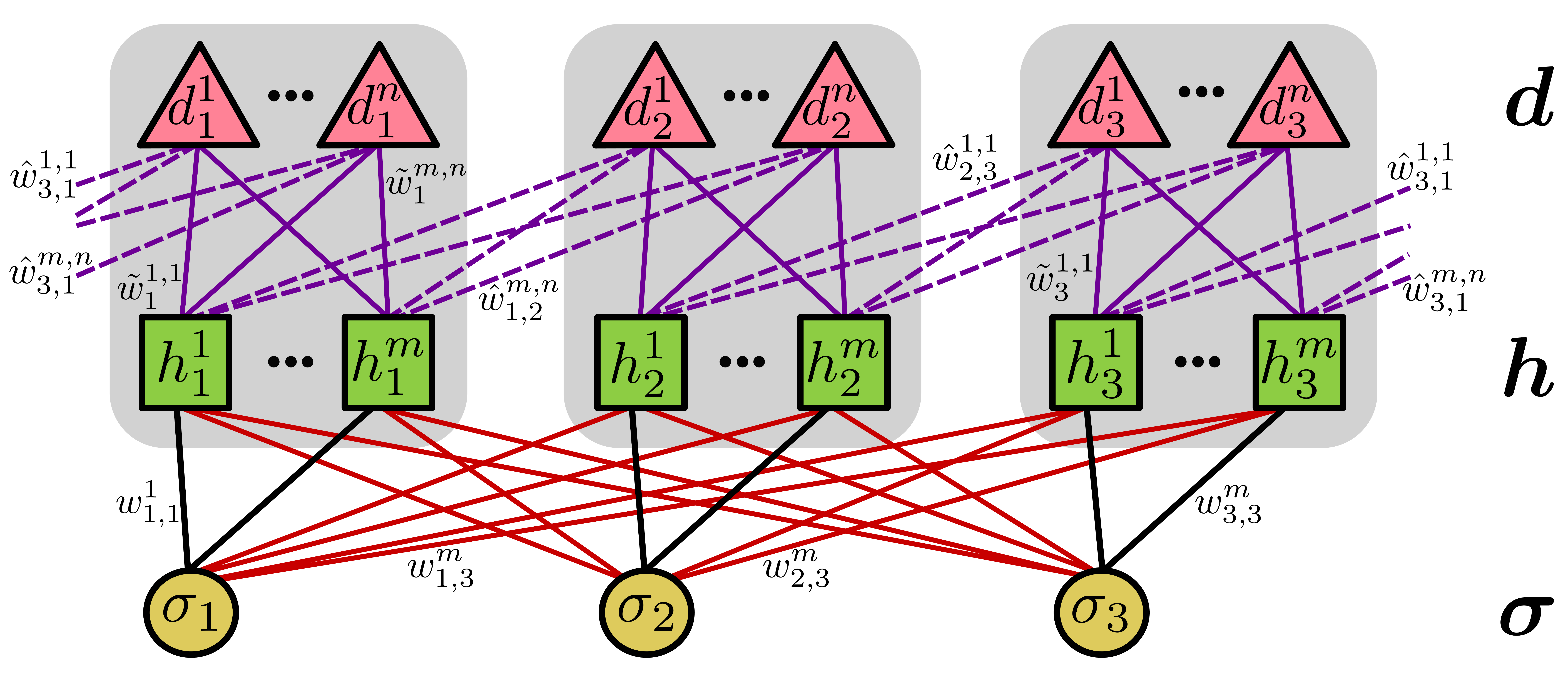}
	\caption{(color online) GTMS network for $N=3$ physical sites. Each transfer matrix is denoted with a grey box. The red links $w_{i,j\ne i}^{\mu}$ are the RBM weights, which result in the dependence of the transfer matrices on all the physical spins. By erasing the RBM links in the architecture one obtains a MPS with bond dimension $2^n$. The purple links are the MPS weights. Bias terms as well as the direct links between sets of deep variables in neighboring blocks are not shown.}
	\label{fig:GTMSfull}
\end{figure*}

\section{DEEP BOLTZMANN MACHINE STATES}
\label{sectio:DBM}
To set the general stage for our construction,  we start with a brief discussion of deep Boltzmann machine (DBM) states \cite{Bengio_RBMrepresentability,HintonDBM,GaoDBM}. We focus on a three-layer architecture, in which the auxiliary units can be organized in two distinct neural layers which we call hidden and deep layers [see Fig \ref{fig:DBM_full}]. The input to such a DBM network is the physical layer, i.e. the set of $N$ physical spins (or any quantum number locally associated to the sites of the physical system) $\sigma_i$ ($i=1,...,N$). The hidden and deep layers are chosen as sets of classical Ising spins with values $h_j=\pm 1$, $d_k=\pm 1$ ($j=1,...,M$, $k=1,...,D$). 
We denote the physical spin configuration with $\boldsymbol{\sigma}=\{\sigma_1,...,\sigma_N\}$, the hidden spin configuration with $\boldsymbol{h}=\{h_1,...,h_M\}$ and the deep spin configuration with $\boldsymbol{d}=\{d_1,...,d_D\}$. The physical and hidden spins are coupled by a set of (complex) weights $w_{i,j}$ representing the links of the network. Hidden and deep layers are coupled as well by a set of weights $\tilde{w}_{j,k}$. Additionally, the weights $c_i$, $b_j$ and $a_k$ are referred to as bias terms, and play the role of local fields for the physical $\sigma_i$, hidden $h_j$ and deep $d_k$ spins, respectively.  All these couplings, collectively denoted by $\boldsymbol{w}=\{c_i,b_j,w_{i,j},\tilde{w}_{j,k},a_k\}_{i,j,k}$ play the role of the variational parameters for the DBM state. The connectivity of the network is encoded in a function, called network energy, which in the present case reads as
\begin{equation}
\begin{split}
\mathcal{E}_{\text{nw}}(\boldsymbol{\sigma},\boldsymbol{h},\boldsymbol{d};\boldsymbol{w})=-\sum_{i=1}^Nc_i\sigma_i+\sum_{k=1}^Da_kd_k \,\,\,\, \\
+\sum_{j=1}^M\big(b_j+\sum_{i=1}^N\sigma_iw_{i,j}+\sum_{k=1}^D\tilde{w}_{j,k}d_k\big)h_j \,\,.
\label{eq:enet}
\end{split}
\end{equation}
The network configurations are then assigned generalized complex Boltzmann weights $\text{e}^{-\mathcal{E}_{\text{nw}}(\boldsymbol{\sigma},\boldsymbol{h},\boldsymbol{d};\boldsymbol{w})}$, and the variational wave function $\psi_{\boldsymbol{w}}(\boldsymbol{\sigma})$ is obtained after a partial partition sum, i.e. by tracing over the hidden and deep layer configurations:
\begin{equation}
\psi_{\boldsymbol{w}}(\boldsymbol{\sigma})=\sum_{\{\boldsymbol{d}\}}\sum_{\{\boldsymbol{h}\}}\text{e}^{-\mathcal{E}_{\text{nw}}(\boldsymbol{\sigma},\boldsymbol{h},\boldsymbol{d};\boldsymbol{w})} \,\,.
\label{eq:partialtrace}
\end{equation}
As a simpler limiting case obtained by discarding the deep layer, let us briefly recall the notion of restricted Boltzmann machine (RBM) states \cite{CarleoTroyer}. In a RBM network the auxiliary spins consist of only one hidden layer of classical Ising spins  $h_j=\pm 1$ and there are no couplings within the set of hidden units. The network energy (see (\ref{eq:enet})) for a RBM simplifies to $\mathcal{E}_{\text{nw}}(\boldsymbol{\sigma},\boldsymbol{h};\boldsymbol{w})=-\sum_ic_i\sigma_i+\sum_j(b_j+\sum_i\sigma_iw_{i,j})h_j$. To obtain $\psi_{\boldsymbol{w}}(\boldsymbol{\sigma})$, for the RBM we only need to sum over all $\boldsymbol{h}$ configurations, which yields the RBM state $\psi_{\boldsymbol{w}}(\boldsymbol{\sigma})=\text{e}^{\sum_ic_i\sigma_i}\prod_{j=1}^{M}2\cosh\big(b_j+\sum_i\sigma_iw_{i,j}\big)$ \cite{CarleoTroyer}.

The power of going from RBM to DBM networks lies in the universal representational capabilities \cite{Bengio_RBMrepresentability} of the latter, which has been demonstrated in a quantum physics context by showing that a three-layer DBM is capable of exactly representing the (imaginary) time evolution of generic quantum many-body systems \cite{Carleo_exactImag}. Concretely, Ref.~\cite{Carleo_exactImag} proves that a suitable DBM network of polynomial complexity in system size and imaginary time $\beta$ with weights $\boldsymbol{w}_{\beta}$ can exactly represent the imaginary time evolution of an initial quantum state $|\psi_0\rangle$ with respect to a generic many-body Hamiltonian $H$, i.e.
\begin{equation}
\psi_{\boldsymbol{w}_{\beta}}(\boldsymbol{\sigma})=\langle\boldsymbol{\sigma}|\text{e}^{-\beta H}|\psi_0\rangle \,\,.
\end{equation}
However, the major caveat limiting the immediate applicability of this strong result is that it is practically impossible in general to exactly evaluate the wave function amplitude $\psi_{\boldsymbol{w}_{\beta}}(\boldsymbol{\sigma})$ by performing the sum on the right-hand side of Eq.~(\ref{eq:partialtrace}). This is in stark contrast to the simpler RBM, where the wave function is readily be calculated analytically. In more physical terms, Boltzmann machine state amplitudes $\psi_{\boldsymbol{w}}(\boldsymbol{\sigma})$ resemble an effective action for the physical spins obtained by tracing out a bath of hidden spin variables. Within this analogy, for the RBM architecture the hidden layer amounts to a free spin-system, while for a DBM the auxiliary variables represent an interacting spin system which is hard to solve in general. In the remainder of this paper we will identify and study a class of DBM architectures, coined generalized transfer matrix state (GTMS) networks, where the axiliary layers can be exactly traced over, and which leads to a unifying generalization of MPS and RBM states.

\section{GENERALIZED TRANSFER MATRIX STATES}
\label{sectio:GTMS}
We now define the central entity of this work, namely the generalized transfer matrix state (GTMS) network, as a particular deep Boltzmann machine which allows for exact traceability of the axiliary layers. Our construction is inspired by the aforementioned interpretation of the DBM wave function as an effective action obtained by tracing out an interacting spin system representing the set of auxiliary units. This raises the natural question what kind of wave functions are obtained when constraining the couplings so as to make this auxiliary spin-system exactly solvable, which leads us to a substantially larger class of networks than the previously considered RBM architecture (corresponding to a non-interacting auxiliary spin system). Specifically, we will group the auxiliary spins into blocks and limit the connectivity between hidden and deep layers to a nearest-neighbor connectivity between these blocks [see grey boxes in Fig.~\ref{fig:generalGTMS}], while retaining all-to-all connectivity between the physical and the hidden layer. Once this constraint has been implemented, the sum over the auxiliary variables can be evaluated adopting a transfer matrix method, well known from the solution of the 1D Ising model.

A detailed exemplary visualization of this GTMS network architecture is shown in Fig.~\ref{fig:GTMSfull}. The hidden and deep auxiliary spins are grouped into $N_T$ blocks (the grey shaded areas in  Figs.~\ref{fig:generalGTMS} and \ref{fig:GTMSfull}), containing $n$ deep and $m$ hidden spins per block. Within these blocks, the connectivity between hidden and deep variables is all-to-all, but to make these auxiliary layers exactly traceable the couplings between different blocks are limited to nearest neighbors [the purple dashed lines in the Fig.~\ref{fig:GTMSfull}]. We impose periodic boundary conditions (PBC) to the network, i.e., the last and the first blocks of auxiliary spins are also coupled. In general the number of blocks $N_T$ can be different from the number of physical sites $N$. Also, arbitrary (i.e. from two-body to $n$-body) direct couplings between deep variables in the same block and in neighboring blocks, as well as direct all-to-all couplings between deep and physical layers, can be introduced, still keeping hidden and deep layers exactly traceable (these connections are not shown in Fig.~\ref{fig:GTMSfull}). Finally, we point out that the number $m$ and $n$ of hidden and deep spins per block can in general depend on the block index itself, i.e. $m=m_{j}$ and $n=n_{j}$ with $j=1,...,N_T$. 

Next, we explicitly perform the sum over hidden and deep units configurations in the GTMS network illustrated in Fig.~\ref{fig:GTMSfull}, so as to derive an analytical form of the GTMS amplitude. For a straightforward extension to the aforementioned slightly more general connectivity we refer to Appendix \ref{sectio:exactlycontractibleGTMS}. The network energy of the GTMS network reads as
\begin{equation}
\begin{split}
\mathcal{E}_{\text{nw}}(\boldsymbol{\sigma},\boldsymbol{h},\boldsymbol{d};\boldsymbol{w})=-\sum_{i=1}^Nc_i\sigma_i+\sum_{j=1}^{N_T}\bigg\{-\sum_{\nu=1}^n a_j^{\nu}d_j^{\nu}\quad\quad\;\;\\
+\sum_{\mu=1}^m \Big[b_j^{\mu}+\sum_{i=1}^N\sigma_iw_{i,j}^{\mu}\quad\quad\quad\quad\;\;\;\;\;\;\\
+\sum_{\nu=1}^n \big(\tilde{w}_{j}^{\mu,\nu}d_j^{\nu}+\hat{w}_{j,j+1}^{\mu,\nu}d_{j+1}^{\nu}\big)\Big]h_j^{\mu}\bigg\} \,\,. \;\;\;
\end{split}
\label{eq:GTMS_netwenergy}
\end{equation}
Here the set of weights $\boldsymbol{w}$ contains: $c_i$, $b_j^{\mu}$, $a_j^{\nu}$ which are the complex on-site bias weights for $\sigma_i$, $h_j^{\mu}$, $d_j^{\nu}$ respectively (not explicitly shown in Fig.~\ref{fig:GTMSfull}), $w_{i,j}^{\mu}$ which denote the couplings between physical $\sigma_i$ and hidden $h_{j}^{\mu}$ (red and purple links between physical and hidden layers in Fig.~\ref{fig:GTMSfull}), $\tilde{w}_{j}^{\mu,\nu}$ that couple $h_{j}^{\mu}$ and $d_j^{\nu}$ within the same block, and $\hat{w}_{j,j+1}^{\mu,\nu}$ that couple $h_{j}^{\mu}$ and $d_{j+1}^{\nu}$ in neighboring blocks (the dashed purple links in Fig.~\ref{fig:GTMSfull}). We refer to the weights $w^{\mu}_{i,j}$ with $i\ne j$ as RBM weights (red links in Fig.~\ref{fig:GTMSfull}), while the rest of the links, except for the $w^{\mu}_{i,i}$ (the black links in Fig.~\ref{fig:GTMSfull}) are referred to as MPS weights. This nomenclature is motivated by the fact that if the network is restricted to contain only MPS weights together with the $w^{\mu}_{i,i}$'s (i.e. if one sets to $0$ the RBM weights), the state obtained from it can be recast as an MPS. This will be explained in more detail in Section IV. If we keep instead only the RBM weights with the $w^{\mu}_{i,i}$'s (that is, we set to $0$ the MPS weights) the dependence of the network energy (Eq.~(\ref{eq:GTMS_netwenergy})) on the deep spins would disappear yielding eventually the network energy of a RBM, and therefore a RBM wave function after the hidden layer configurations have been summed over.

To perform the sum over hidden and deep variables configurations of Eq.~(\ref{eq:partialtrace}) we first trace out the hidden layer. Since the hidden spins are not directly coupled, this sum is easily performed (analogous to the RBM case), and yields
\begin{equation}
\psi_{\boldsymbol{w}}(\boldsymbol{\sigma})=\text{e}^{\sum_ic_i\sigma_i}\sum_{\{\boldsymbol{d}\}}\prod_jt_{j,j+1}(\boldsymbol{\sigma},\boldsymbol{d}_j,\boldsymbol{d}_{j+1}) \,\,,
\label{eq:psiGTMS_predeeptrace}
\end{equation}
where $\boldsymbol{d}_j$ denotes the deep spin configuration at block $j$, i.e. $\boldsymbol{d}_j=\{d_j^{\nu}\}_{\nu=1,...,n}$. The elements $t_{j,j+1}$ of the product read as
\begin{equation}
t_{j,j+1}(\boldsymbol{\sigma},\boldsymbol{d}_j,\boldsymbol{d}_{j+1})=\text{e}^{\sum_{\nu}a_j^{\nu}d_j^{\nu}}\prod_{\mu=1}^m 2\cosh\big(\varphi_j^{\mu}(\boldsymbol{\sigma},\boldsymbol{d}_j,\boldsymbol{d}_{j+1})\big) \,\,,
\label{eq:GTMS_t_product_elements}
\end{equation}
with
\begin{equation}
\begin{split}
\varphi_j^{\mu}(\boldsymbol{\sigma},\boldsymbol{d}_j,\boldsymbol{d}_{j+1})=b_j^{\mu}+\sum_{i=1}^N\sigma_iw_{i,j}^{\mu}\quad\quad\quad\quad\;\;\\
+\sum_{\nu=1}^n \big(\tilde{w}_{j}^{\mu,\nu}d_j^{\nu}+\hat{w}_{j,j+1}^{\mu,\nu}d_{j+1}^{\nu}\big) \,\,.
\end{split}
\label{eq:phiangles_GTMS}
\end{equation}
Now, in order to perform the sum over all the configurations of the deep variables, we can interpret the complex numbers $t_{j,j+1}(\boldsymbol{\sigma},\boldsymbol{d}_j,\boldsymbol{d}_{j+1})$ as elements of a \emph{transfer matrix} $T_j(\boldsymbol{\sigma})$ associated to block $j$. One can uniquely associate a index, running from $1$ to $2^n$, to a deep spin configuration $\boldsymbol{d}_j$ at block $j$, by interpreting the $n$ Ising spins $d_j^{\nu}$ in $\boldsymbol{d}_j$ as bits. We can define the elements of the transfer matrix as
\begin{equation}
\Big(T_j(\boldsymbol{\sigma})\Big)_{\boldsymbol{d}_j,\boldsymbol{d}_{j+1}}=t_{j,j+1}(\boldsymbol{\sigma},\boldsymbol{d}_j,\boldsymbol{d}_{j+1}) \,\,.
\label{eq:deftransfer}
\end{equation}
Tracing out the deep layer in Eq.~(\ref{eq:psiGTMS_predeeptrace}) is then equivalent to taking the product of these transfer matrices, i.e.,
\begin{equation}
\psi_{\boldsymbol{w}}(\boldsymbol{\sigma})=\text{e}^{\sum_ic_i\sigma_i}\,\text{tr}\bigg(\prod_{j=1}^{N_T}T_j(\boldsymbol{\sigma})\bigg), 
\label{eq:GTMS}
\end{equation}
where the trace comes from the periodic boundary conditions imposed on the blocks of auxiliary spins of the network. We stress that introducing PBC in the GTMS network is not a necessary requirement for the exact traceability of the auxiliary layers. A GTMS network without PBC can equally well be used for physical systems with open boundary conditions (see Appendix \ref{sectio:GTMSwithOBC} for a more detailed discussion).

In case $m=m_{j}$ and $n=n_{j}$ then the dimension of $T_j$ would depend on $j$ as well, being equal to $2^{n_j}\times 2^{n_{j+1}}$. The transfer matrices depend in general on the index $j$ as well as on the physical spin configuration $\boldsymbol{\sigma}$ over the entire system, as opposed to the well known case of MPS where each matrix depends locally on the spin quantum number on one physical site. For this reason we call the state of Eq.~(\ref{eq:GTMS}) a generalized transfer matrix state (GTMS), and we will show below that this nonlocal dependence on the physical quantum numbers allows the GTMS to capture long-range correlations going beyond the area-law typical of MPS.

\section{MATRIX PRODUCT STATES FROM GTMS}
\label{sectio:MPSfromGTMS}
In this section, we demonstrate how, by removing the RBM weights, the GTMS network is able to parametrize generic MPS, with a bond dimension $2^n$ set by the number $n$ of deep spins per physical site. Defined as a product of tensors with elements $A^{[i]\sigma_i}_{a_{i-1}a_i}$ associated to each physical site $i$ with one physical index $\sigma_i$ and two auxiliary indices $a_1,a_2=1,...,\chi$ with the bond dimension $\chi$, a generic MPS is of the form \cite{SchollwoeckMPS_DMRG,Cirac_MPS_DMRG,EisertPlenio_AreaLaw}
\begin{equation}
\psi_{\text{MPS}}(\boldsymbol{\sigma})=\text{tr}\bigg(\prod_{i=1}^{N}A^{[i]\sigma_i}\bigg) \,\,.
\label{eq:MPS_PBC}
\end{equation}
We can immediately see that this form is similar to the one of Eq.~(\ref{eq:GTMS}) with the number of transfer matrices $N_T$ equal to the number of physical sites $N$, apart form the fact that here the matrices in the product depend only on the quantum number $\sigma_i$ of the physical site $i$. To reduce Eq.~(\ref{eq:GTMS}) to the MPS form of Eq.~(\ref{eq:MPS_PBC}) we simply restrict the connectivity of the GTMS network so as to make the $T_i$ depend only on $\sigma_i$. We note that in Eq.~(\ref{eq:GTMS}) the dependence of $T_j$ on the entire physical layer enters via the angles $\varphi_j^{\mu}$ of Eq.~(\ref{eq:phiangles_GTMS}), where the term $\sum_i\sigma_iw_{i,j}^{\mu}$ appears. Therefore, if we set to $0$ all the couplings $w_{i,j}^{\mu}$ where $i\ne j$ each $\varphi_j^{\mu}$ depends only on $\sigma_j$. Pictorially this amounts to erasing all the red links in Fig.~\ref{fig:GTMSfull}. In physical terms, this corresponds to neglecting the long-range quantum correlations which are mediated by the RBM couplings, keeping only the short-range correlations encoded in the MPS couplings between neighboring transfer matrices. This way, Eq.~(\ref{eq:GTMS}) becomes formally analogous to the MPS in Eq.~(\ref{eq:MPS_PBC}), with bond dimension $\chi=2^n$ (assuming $n$ constant throughout the system and $N_T=N$):
\begin{equation}
\psi_{\boldsymbol{w}}(\boldsymbol{\sigma})=\text{tr}\bigg(\prod_{i=1}^{N}\text{e}^{c_i\sigma_i}T_i(\sigma_i)\bigg) \,\,,
\label{eq:MPS_fromGTMS}
\end{equation}
where $\text{e}^{c_i\sigma_i}T_i(\sigma_i)$ can be identified with the tensor $A^{[i]\sigma_i}$ in Eq.~(\ref{eq:MPS_PBC}) and the notation $T_i(\sigma_i)\equiv T_i(\boldsymbol{\sigma})$ has been introduced to make the local dependence of the transfer matrices on the physical spins manifest.

\begin{figure} 
	\includegraphics[width=0.95\columnwidth]{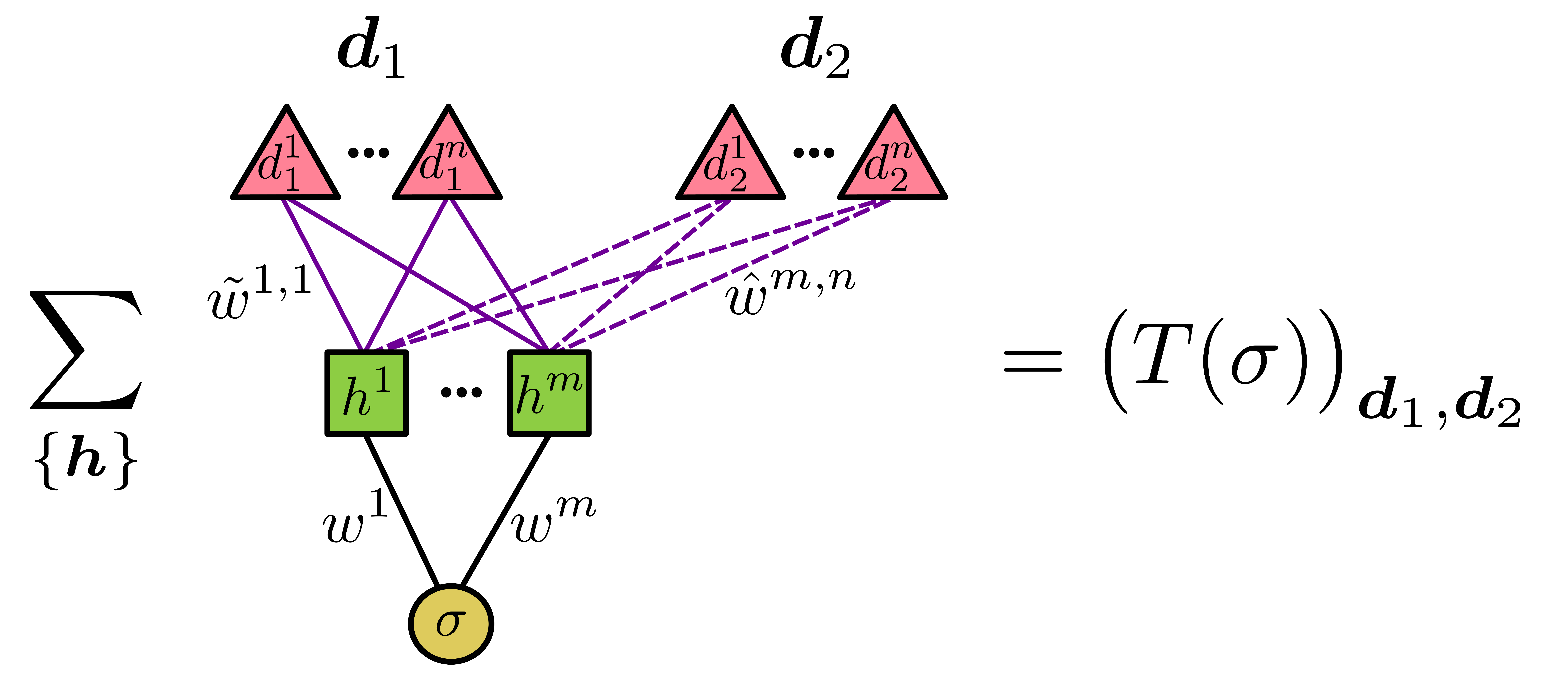}\\
	\caption{(color online) Block of GTMS network representing the elements $\big(T(\sigma)\big)_{\boldsymbol{d}_1,\boldsymbol{d}_2}$ of a MPS tensor with bond dimension $\chi=2^n$ and physical index $\sigma$. A single matrix element is obtainend by fixing the deep spin configurations $\boldsymbol{d}_1,\boldsymbol{d}_2$ and $\sigma$, and then computing the trace over the hidden spin configurations $\boldsymbol{h}=\{h_1,...,h_m\}$.}
	\label{fig:GTMS_MPStensor}
\end{figure}

We note that for being able to parametrize an arbitrary MPS with bond dimension $2^n$ and spin $s$ physical degrees of freedom, one would in general need $2^{2n}(2s+1)$ complex free parameters per MPS tensor. Simple parameter counting shows that, in order to have enough free parameters, the number $m$ of hidden spins per block should scale with $n$ according to $m\approx\frac{(2s+1)2^{2n}}{2(n+1)}$, i.e. quadratically with the bond dimension of the MPS. This can be seen as a consequence of the universal approximator property of RBMs \cite{Bengio_RBMrepresentability} and its extension to the complex case \cite{HuangPiP}. To this end we interpret the $2n$ deep variables of a GTMS block representing a single MPS tensor (see Fig. \ref{fig:GTMS_MPStensor}) as a the visible layer of a RBM where the $m$ hidden variables are to be traced out, and the physical spin represents a fixed parameter not to be summed over. The aforementioned approximation theorems \cite{Bengio_RBMrepresentability,HuangPiP} applied to this construction then predict that with a number of hidden variables $m\approx 2^{2n}$ any vector in $\mathbb{C}(2^{2n})$, that is any MPS tensor, can be represented. This will be also confirmed later by some numerical examples.\newline We also note in passing that given the ability of our GTMS ansatz to parametrize arbitrary MPS, one can easily construct an higher-dimensional generalization of the GTMS network which can parametrize generic string-bond states (SBS) \cite{Schuch_Cirac_SBS} (see Appendix \ref{sectio:SBS} for further details). Importantly, in such GTMS-SBS generalization, the property of exact traceability of hidden and deep layers is still preserved. Thus our GTMS approach is in principle not only limited to one dimensional systems. In principle a higher-dimensional analog of GTMS may also be based on higher-dimensional tensor network states such as PEPS \cite{CiracReview_TNS}. However, in this case, similar to the PEPS ansatz itself, approximate schemes for the calculation of the state amplitude from such a GTMS generalization would be required.

\subsection{AKLT State from GTMS}
\begin{figure} 
	\includegraphics[width=0.95\columnwidth]{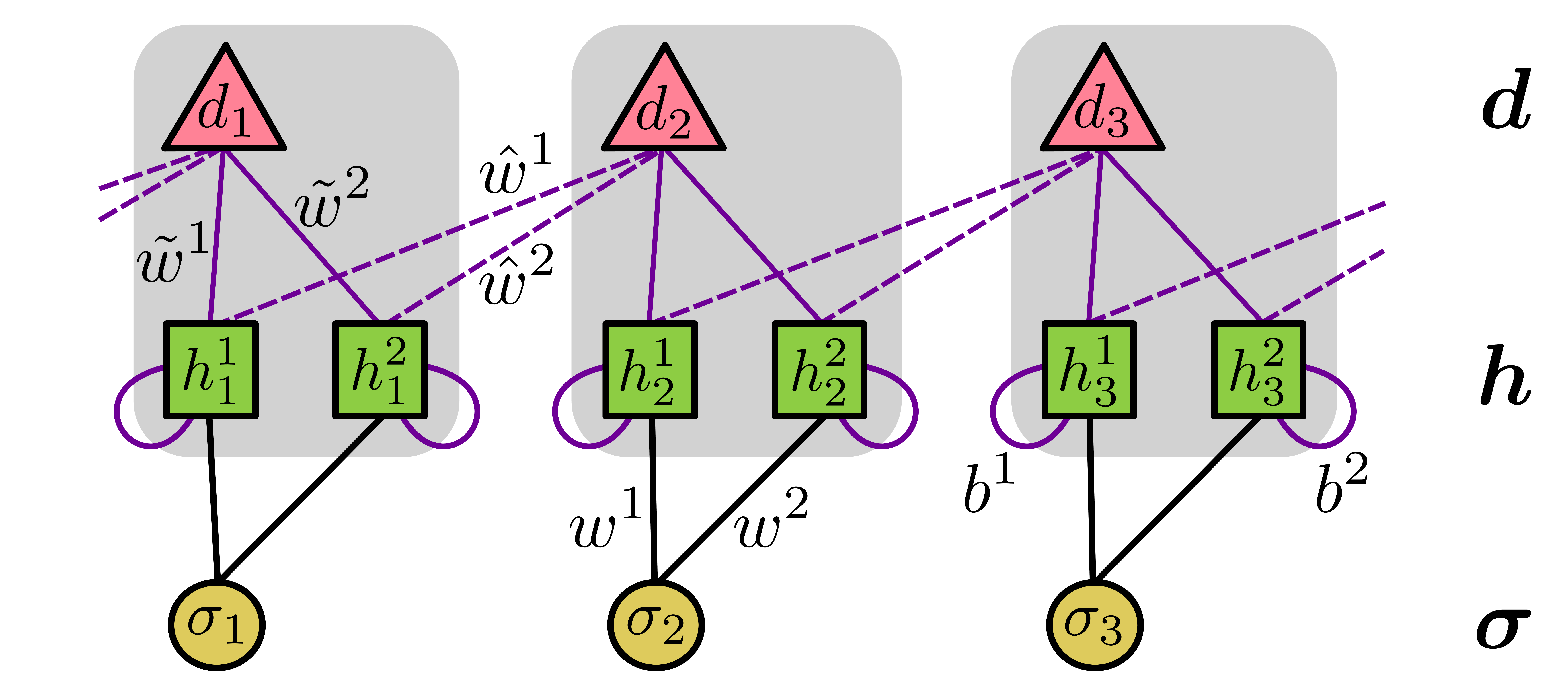}\\
	\caption{(color online) GTMS network for the construction of the AKLT state, for $N=3$ physical sites, with the bias terms for hidden units explicitly shown. Since the AKLT state is translation invariant we have that $\forall\,i=1,...,N$ $w_i^{\mu}=w^{\mu}$, $\tilde{w}_i^{\mu}=\tilde{w}^{\mu}$, $\hat{w}_{i,i+1}^{\mu}=\hat{w}^{\mu}$, $b_i^{\mu}=b^{\mu}$, with $\mu=1,2$.}
	\label{fig:GTMS_AKLT}
\end{figure}

As an emblematic example we explicitly construct the AKLT state \cite{AKLTpaper,AKLTpaper2} from a GTMS network, shown in Fig.~\ref{fig:GTMS_AKLT}. The AKLT state is one of the simplest MPS with bond dimension $2$ and tensors independent of position. This suggests that the use of a GTMS architecture with constant $n=1$, and $m=2$ hidden variables per site will be sufficient for fully parametrizing the state. The AKLT state is the ground state of a modified spin-1 quantum Heisenberg model \cite{AKLTpaper}, hence the physical spin variables which constitute the inputs of our network can take values $\sigma_i=-1,0,+1$ (while in the remainder of the paper we will use spin $1/2$ degrees of freedom). The normalized AKLT matrices read
\begin{equation}
\begin{split}
A^+=\frac{2}{\sqrt{3}}
\begin{pmatrix}
0 & \frac{1}{\sqrt{2}} \\
0 & 0
\end{pmatrix}, \quad\quad
A^-=\frac{2}{\sqrt{3}}
\begin{pmatrix}
0 & 0 \\
- \frac{1}{\sqrt{2}} & 0
\end{pmatrix}, \\
A^0=\frac{2}{\sqrt{3}}
\begin{pmatrix}
-\frac{1}{2} & 0 \\
0 & \frac{1}{2} 
\end{pmatrix} \,\,. \quad\quad\quad\quad\quad
\end{split}
\label{eq:AKLT_matrices}
\end{equation}
For the ANN architecture shown in Fig.~\ref{fig:GTMS_AKLT}, we obtain the $2\times2$ transfer matrices
\begin{equation}
\big(T(\sigma_i)\big)_{d_i,d_{i+1}}=\prod_{\mu=1}^{2}\cosh\big(\varphi^{\mu}(\sigma_i,d_i,d_{i+1})\big) \,\,,
\end{equation}
with the angular arguments
\begin{equation}
\varphi^{\mu}(\sigma_i,d_i,d_{i+1})=b^{\mu}+\sigma_iw^{\mu}+\tilde{w}^{\mu}d_i+\hat{w}^{\mu}d_{i+1} \,\,,
\end{equation}
where the independence of several quantities on the physical site index $i$ reflects the translation invariance of the AKLT state.
By choosing $w^{1}=i\pi/4$, $w^{2}=i\pi/2$, $\tilde{w}^{1}=-i\pi/4$, $\tilde{w}^{2}=i3\pi/4$, $\hat{w}^{1}=0$, $\hat{w}^{2}=i3\pi/4$, $b^{1}=0$ and $b^{2}=-i\pi/2$, we find $\sqrt{2/3}\,T(\sigma_i)=A^{\sigma_i}$, where the normalization factor $\sqrt{2/3}$ can formally be added to the DBM network by a constant shift to the network energy.

This  explicit parametrization demonstrates how the AKLT state is exactly represented by a \emph{short range} DBM network, in which the connectivity is limited to neighboring blocks of auxiliary units. Interestingly, this simple state \emph{cannot} be directly represented by a short-range RBM network. Indeed, as noticed in Ref.~\cite{GlasserMunich} short-range RBM states correspond to so called entangled-plaquette states, which are products of complex numbers associated to local clusters (plaquettes) of physical sites. Physically, this product structure of commuting local factors makes it impossible for such states to encode the hidden infinitely-ranged string order of the AKLT state \cite{AKLTstringorder}. In more practical terms, the string order constrains the AKLT wave function to vanish whenever two subsequent $A^+$ or two subsequent $A^-$ matrices at the physical sites $i$ and $j$ are separated only by $A^0$ matrices, no matter how large the distance between $i$ and $j$ is. This is encoded in the basic (non-commuting) algebra of these matrices. Clearly, such a constraint  cannot be achieved by a product of complex numbers that depend only locally on the physical variables, as in the case of a short-range RBM.

\subsection{Learning random MPS}
Generalizing from the basic example of the AKLT state, we now confirm numerically that GTMS networks without RBM weights (red links in Figs.~\ref{fig:DBM_GTMS} and \ref{fig:GTMSfull}) can {\emph{learn}} a generic MPS. In the context of artificial neural networks, the word \emph{learning} means that the weights of the network are iteratively optimized to find the minimum of a certain cost function \cite{LeCun_DeepLearning,CarasquillaMelko,Wang_LearningPhases,Bottou_MLwithSGD,Bottou_Optimization_in_ML}. Assuming translation invariance, we optimize only for one MPS tensor with a network (corresponding to a single grey shaded block in Fig.~\ref{fig:GTMSfull}), containing $n$ deep variables per physical site, thus yielding a bond dimension $\chi=2^n$. The resulting network representation of the MPS tensor elements is visualized in Fig.~\ref{fig:GTMS_MPStensor}. By fixing $\sigma$ and the deep spin configurations $\boldsymbol{d}_1=\{d_1^1,...,d_1^n\}$ and $\boldsymbol{d}_2=\{d_2^1,...,d_2^n\}$ one obtains, after tracing out the hidden layer in the network of Fig.~\ref{fig:GTMS_MPStensor}, the $(\boldsymbol{d}_1,\boldsymbol{d}_2)$ element of the transfer matrix $T(\sigma_i)$. 

The cost function $D_\text{rel}$ that we optimize for is defined using the Frobenius norm of the difference between the GTMS transfer matrix $T$ and the random MPS tensor $A$ to be learned:
\begin{equation}
D_\text{rel}\equiv \lVert T-A\rVert_{\text{rel}}^2=\frac{\lVert T-A\rVert^2}{\lVert A\rVert^2} \,\,,
\end{equation}
where the dependence on the variational parameters $\boldsymbol{w}$ lies in the transfer matrix $T$ (see Eqs.~(\ref{eq:GTMS_t_product_elements}-\ref{eq:deftransfer})).

\begin{figure} 
	\includegraphics[width=0.95\columnwidth]{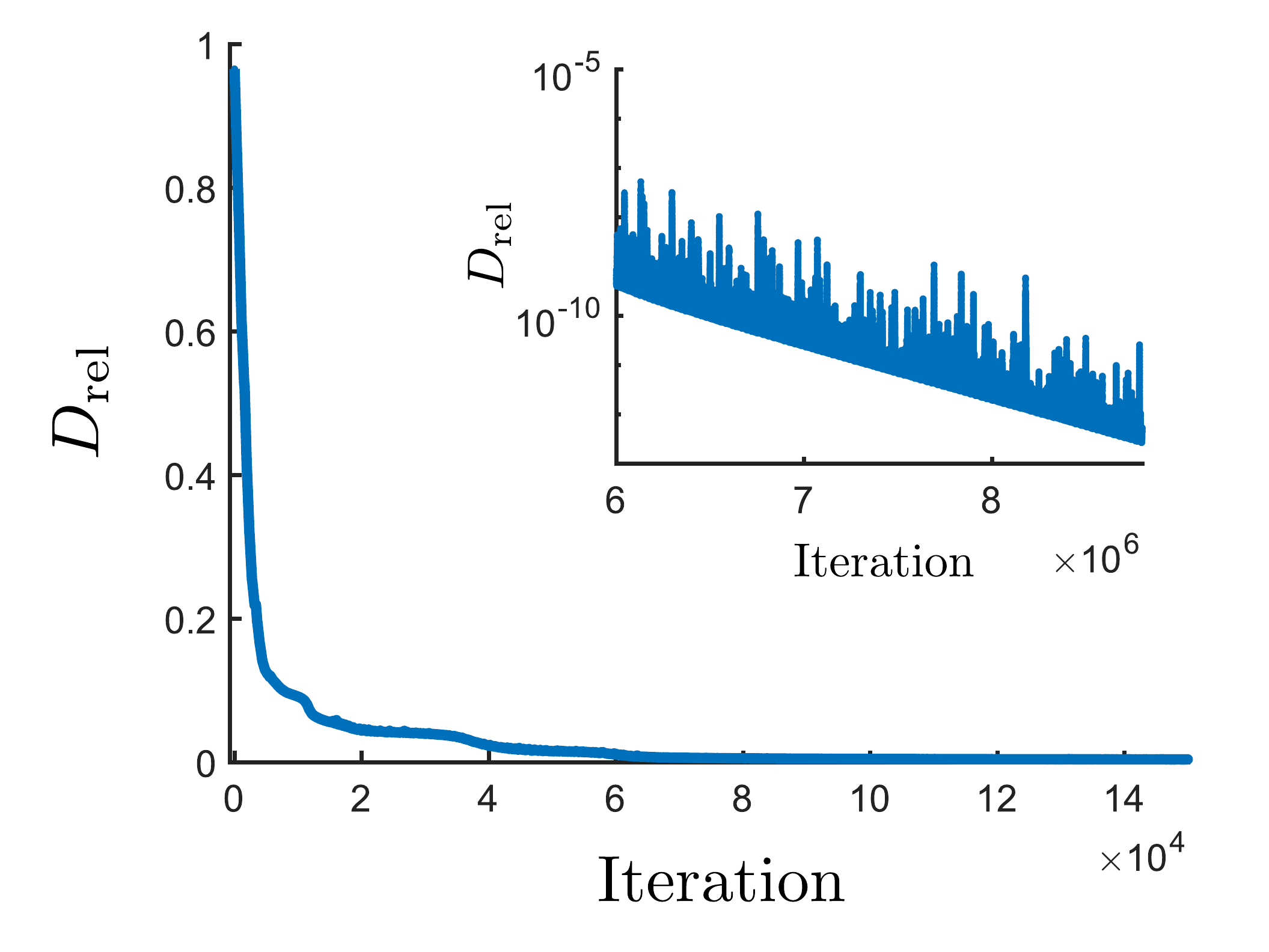}\\
	\caption{(color online) Convergence plot for the learning of a random MPS tensor with bond dimension $\chi=8$ and local Hilbert space dimension $2$, with a network having $n=3$, $m=16$. Here the optimization was done using AdaGrad method. The inset shows a closeup of the cost function at later iterations, plotted in logarithmic scale.}
	\label{fig:learningMPS_bond8}
\end{figure}

We considered the case of spin-$1/2$ degrees of freedom per physical site, so the total number of elements of the MPS tensor is $N_{\text{el}}=2 \chi^2$. The optimization has been performed using stochastic gradient descent methods \cite{Bottou_MLwithSGD,Bottou_Optimization_in_ML} such as AdaGrad and Adam \cite{Duchi_AdaGrad,Adam} with $N_w = 1+2m+n+2mn \simeq 2 \chi^2$ parameters. An example of a convergence plot for $\chi=8$ using AdaGrad for optimizing the network with $n=3$ and $m=16$ is given in Fig.~\ref{fig:learningMPS_bond8}. By using Adam optimizer implemented in the Phyton TensorFlow libraries we were able to learn random MPS tensors up to bond dimension $\chi=16$ to a final relative accuracy $D_\text{rel}\sim 10^{-4}$, using $n=4$ and $m=52$ on an ordinary desktop computer.

\begin{figure} 
	\includegraphics[width=0.95\columnwidth]{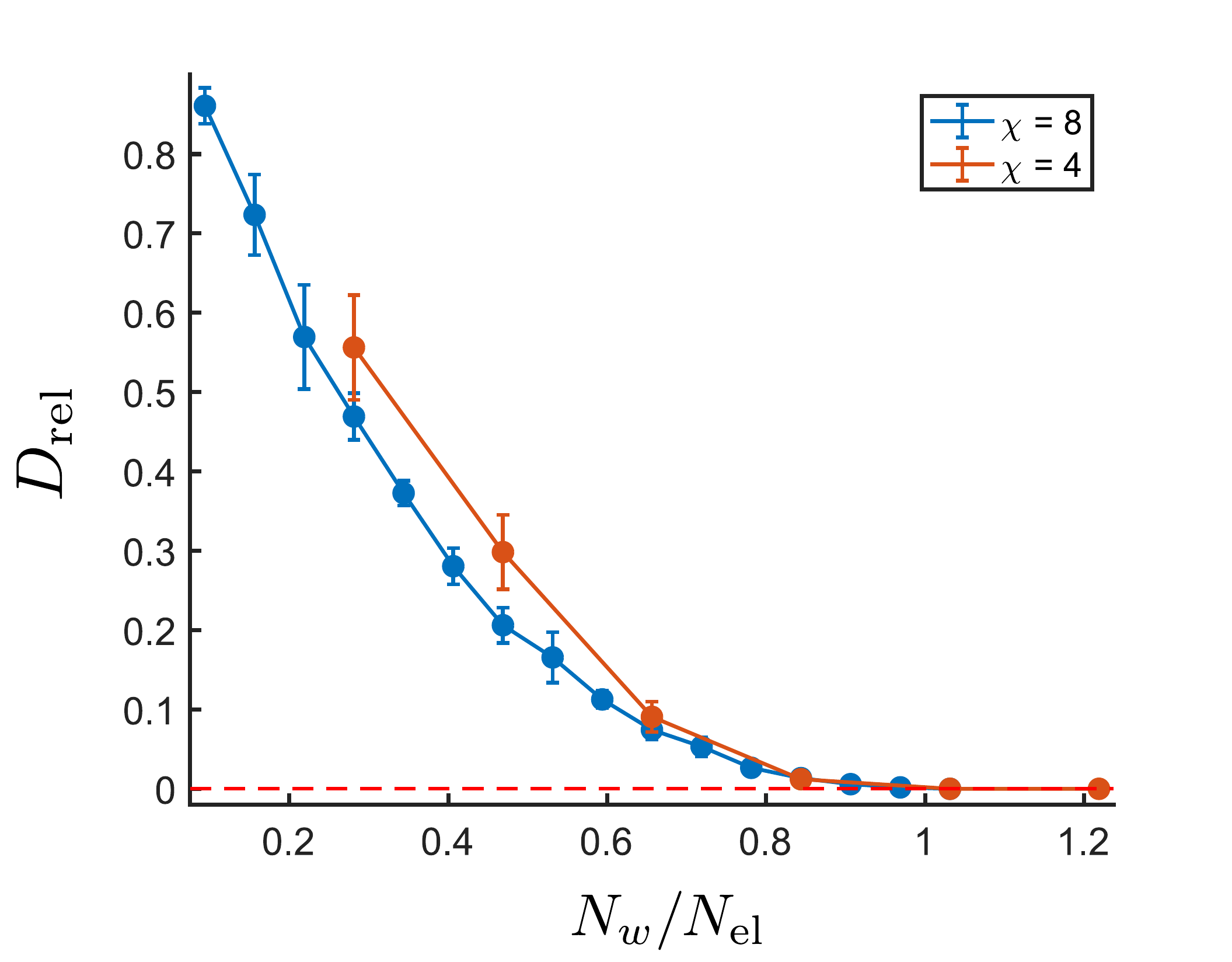}\\
	\caption{(color online) Attempted MPS compression for bond dimension $\chi=4$ (orange dots) and $\chi=8$ (blue dots), performed by removing hidden spins from the network to be optimized. The converged value of $D_\text{rel}$, averaged over ten random MPS tensor realizations, is plotted vs. the ratio $N_w/N_{\text{el}}$. The dashed red line indicates the position of the $0$, for guiding the eye.}
	\label{fig:MPScompression}
\end{figure}

At last we would like to address the question whether it is possible to learn a MPS tensor with a number $N_w$ of variational parameters that is lower than the number of tensor elements $N_{\text{el}}$, thus attempting an approximate compression of the MPS. The results of this compression are shown in Fig.~\ref{fig:MPScompression} relative to a set of ten realizations of random MPS tensors with bond dimension $\chi=4$ (orange data) and $\chi=8$ (blue data). We can see that for $N_w\ge N_{\text{el}}$ our network can indeed be optimized to learn the MPS tensors (each data point in Fig.~\ref{fig:MPScompression} represents the best achieved relative accuracy $\langle D_\text{rel}\rangle$ averaged over the ten random MPS tensor realizations). As $N_w<N_{\text{el}}$, we find however $\langle D_\text{rel}\rangle>0$, meaning that an efficient network compression in general is not viable when learning generic (random) MPS tensors.

\floatsetup[figure]{style=plain,subcapbesideposition=top}
\begin{figure} [htp]
	\sidesubfloat[]{\includegraphics[width=0.95\columnwidth]{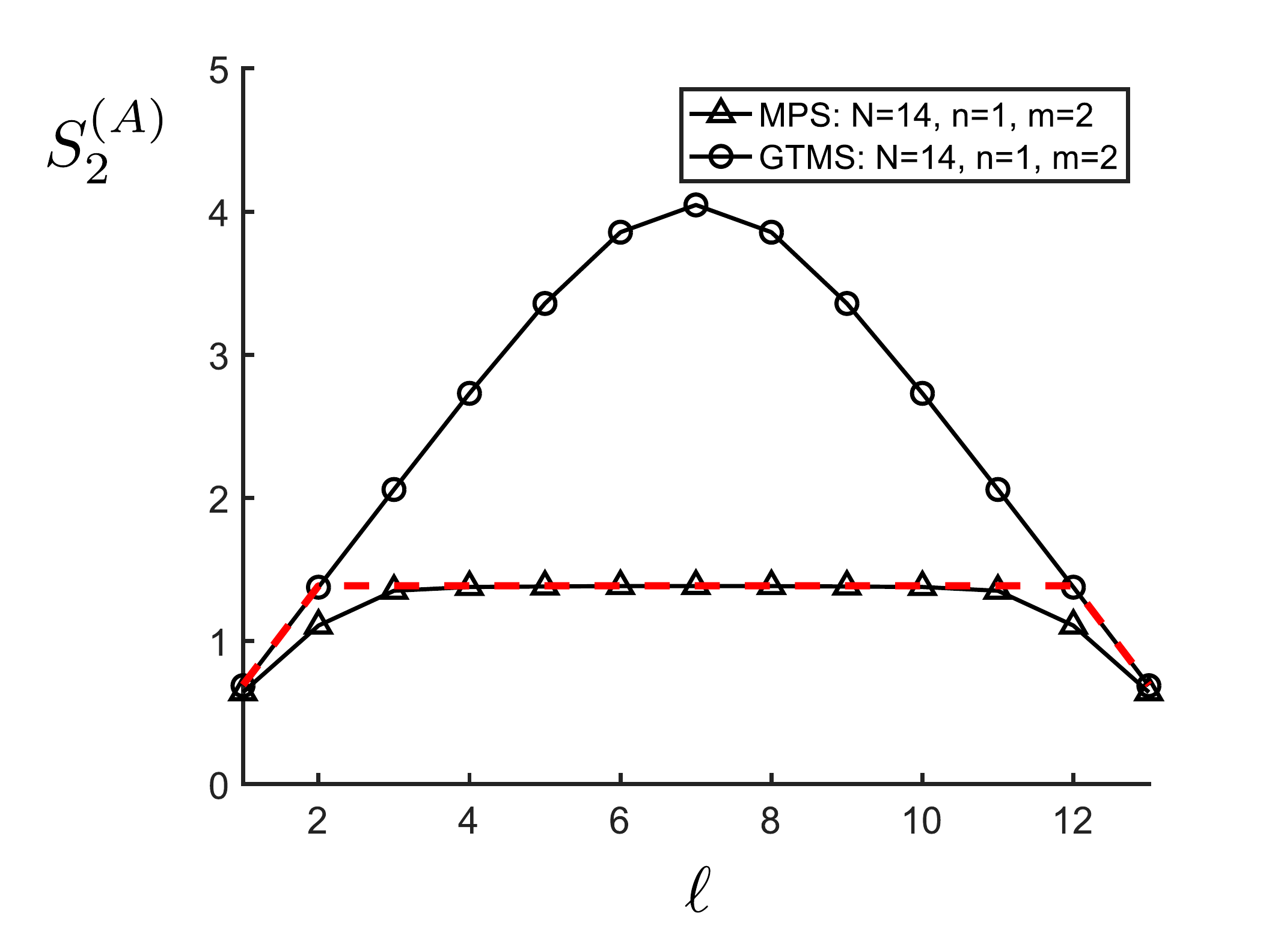}\label{fig:entro_L14_n1m2_exact}}\\ \sidesubfloat[]{\includegraphics[width=0.95\columnwidth]{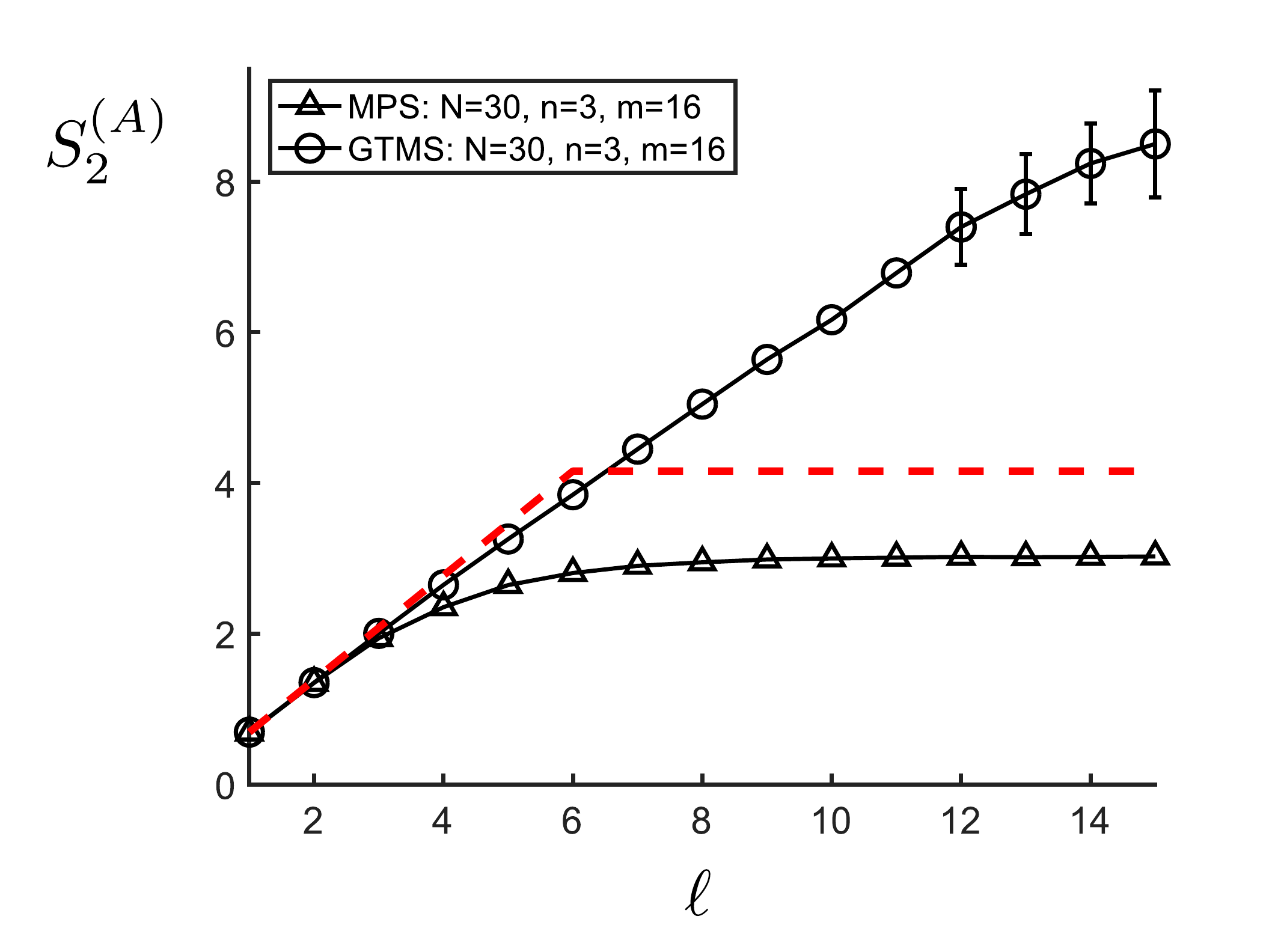}\label{fig:entro_L30_n3m16_sampling}}\\
	\caption{(color online) $S^{(A)}_2$ as a function of the length $\ell$ of subsystem $A$. Black triangles show $S^{(A)}_2$ for a translation invariant MPS obtained from a GTMS network with randomly chosen weights. Black circles are for GTMS obtained by adding random nonlocal RBM couplings to the network parametrizing the MPS. The dashed red line shows the upper bound for the entanglement entropy of a MPS with $\chi=2^n$. Weights are drawn within uniform distribution around $0$ with width $0.2$ and $8\pi$ for real and imaginary parts, respectively. (a) Exact results for $n=1$, $m=2$ and $N=14$ sites. (b) Monte Carlo results for $n=3$, $m=16$ with $N=30$ sites. The statistical error is smaller than the data symbol, when the errorbar is not shown. For translation invariance, we show $S_2^{(A)}$ only up to half of system size.}
	\label{fig:entro_L14}
\end{figure}
\section{ENTANGLEMENT ANALYSIS OF GTMS}
\label{sectio:entanglementGTMS}
In this section, we present a numerical analysis of the entanglement entropy of GTMS for one dimensional systems with PBC and spin-$1/2$ degrees of freedom per site. We will show that the addition of nonlocal RBM weights to a GTMS network representing a MPS results in the onset of volume-law entanglement, as opposed to the area-law scaling obtained when keeping the MPS weights only. This is a clear indication of the improved representational power of generalized transfer matrix states.

Specifically, we calculate the second R\'enyi entropy $S^{(A)}_2$ for a bipartition of a one dimensional spin-$1/2$ system with PBC in two subsystems $A$ and $B$, with the total system being in the pure state $|\psi_{\boldsymbol{w}}\rangle$ with GTMS wave functions of Eq.~(\ref{eq:GTMS}). The second R\'enyi entropy of subsystem $A$ is given by
\begin{equation}
S^{(A)}_2=-\ln\big(\text{tr}\,\rho_A^2\big) \,\,,
\end{equation}
with $\rho_A=\text{tr}_B\,|\psi_{\boldsymbol{w}}\rangle\langle\psi_{\boldsymbol{w}}|$ being the reduced density matrix of subsystem $A$. The algorithm introduced in Ref.~\cite{Hastings_RenyiEntropy} offers a simple and efficient way for calculating $S^{(A)}_2$ with Monte Carlo, which requires Metropolis sampling of two copies of the system, as the trace of $\rho_A^2$ needs to be evaluated. For the $\alpha^{\text{th}}$ R\'enyi entropy one would need to sample configurations of $\alpha$ copies  (see Ref.~\cite{Hastings_RenyiEntropy} and Appendix \ref{sectio:RenyiEntropyVMC} for more details).

We determine the scaling of the second R\'enyi entropy with the length $\ell$ of subsystem $A$, comparing the two cases of a GTMS network parametrizing a conventional MPS, and the augmented GTMS to which nonlocal RBM couplings have been added while keeping the existing couplings unchanged.
In Fig.~\ref{fig:entro_L14}, we show exact data on $S^{(A)}_2$ for a system of $N=14$ sites in the case of a GTMS with $n=1$, $m=2$ (panel \ref{fig:entro_L14_n1m2_exact}), and Monte Carlo data from Metropolis sampling for a system of $N=30$ sites in the case of a GTMS with $n=3$, $m=16$ (panel \ref{fig:entro_L30_n3m16_sampling}). We observe that the addition of RBM couplings results in a volume-law (i.e. linear in $\ell$) scaling of the entanglement, and that $S^{(A)}_2$ exceeds the MPS bound $2\ln\chi$ (the dashed red line in Fig. \ref{fig:entro_L14}) which is set by the bond dimension $\chi=2^n$. In this sense, the GTMS family combines the properties of conventional MPS and RBM states.

\begin{figure}  [htp]
	\includegraphics[width=0.95\columnwidth]{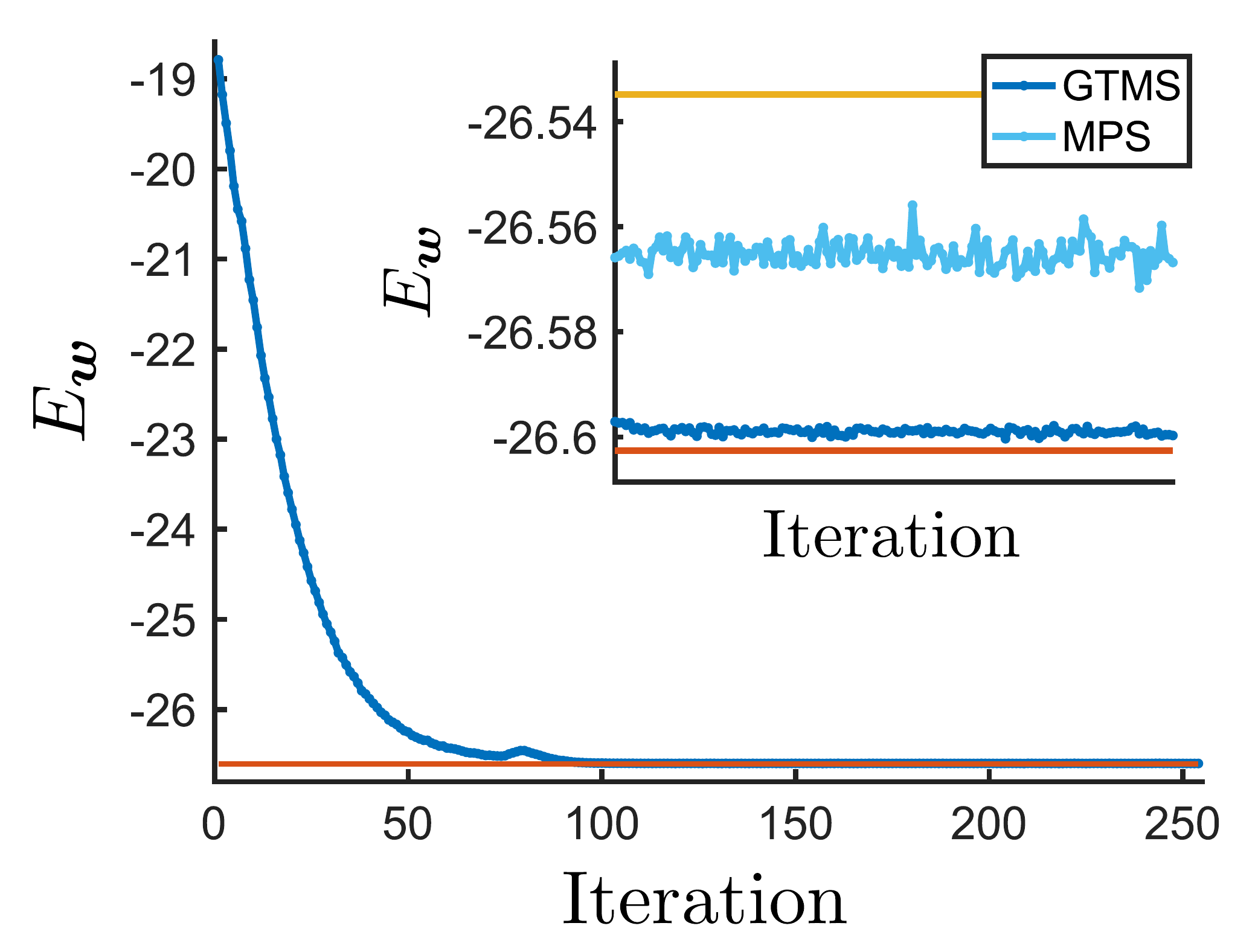}\\
	\caption{(color online) Energy expectation value $E_{\boldsymbol{w}}$ as a function of the SR iterations, for a chain with $N=60$ sites, governed by the Hamiltonian of Eq. (\ref{eq:XXZhamiltonian}). The blue line in the main plot and in the inset shows $E_{\boldsymbol{w}}$ for a translation invariant GTMS ansatz with $n=2$ deep and $m=4$ hidden units per physical site, for a total of $263$ variational parameters. The light blue line in the inset shows $E_{\boldsymbol{w}}$ for a translation invariant MPS with $n=2$ deep and $m=4$ hidden units per tensor. The solid orange line marks the exact value of the ground-state energy $E_{_{\text{GS}}}$ obtained with DMRG (using MPS with open boundaries and a bond dimension of $\chi=2000$, for which the truncation error was $10^{-12}$), and the solid yellow line in the inset marks the exact value of the energy of the first excited state on the zero-magnetization sector, obtained from finite-size scaling of ED results.}
	\label{fig:energyevoGTMS_L60}
\end{figure}

\section{CRITICAL GROUND STATES WITH GTMS}
\label{sectio:XXZ}
In this section, we explicitly demonstrate through a numerical variational Monte Carlo (VMC) calculation that the GTMS ansatz is indeed able to parametrize the ground state of a critical quantum many-body system. The model chosen for our example is the one dimensional spin-1/2 XXZ chain with PBC, described by the following Hamiltonian:
\begin{equation}
H=-J\sum_{j=1}^N\big(S^x_jS^x_{j+1}+S^y_jS^y_{j+1}\big)+\Delta\sum_{j=1}^N S^z_jS^z_{j+1} \,\,,
\label{eq:XXZhamiltonian}
\end{equation}
where the $S^{\alpha}_j$ ($\alpha=x,y,z$) are the spin-1/2 operators fulfilling the standard angular momentum algebra $\big[S^{\alpha}_j,S^{\beta}_{j'}\big]=i\,\delta_{j,j'}\epsilon^{\alpha\beta\gamma}S^{\gamma}_j$ (we set $\hbar=1$), and $N+1\equiv 1$ for PBC. In the simulations we concentrate on the regime $J=1$ and $\Delta=1$, in which the Hamiltonian (\ref{eq:XXZhamiltonian}) is critical and exactly at the Kosterlitz-Thouless transition point. Because of PBC we implement translation invariance in our GTMS network in order to reduce the number of variational parameters (see Appendix \ref{sectio:translinvGTMS} for more details). Furthermore, we restrict the ground state search in the zero-magnetization sector.

We adopt the stochastic reconfiguration (SR) method \cite{CasulaSorellaSR,SorellaSR} for the optimization of the variational parameters $\boldsymbol{w}$, in order to reach the minimum value of the energy expectation value $E_{\boldsymbol{w}}=\frac{\langle\psi_{\boldsymbol{w}}|H|\psi_{\boldsymbol{w}}\rangle}{\langle\psi_{\boldsymbol{w}}|\psi_{\boldsymbol{w}}\rangle}$. We refer the reader to Appendix \ref{sectio:StochReconf} and \cite{CarleoTroyer,our_article} for more details on the practical implementation. We start the VMC ground-state search by randomly initializing the complex variational weights with real and imaginary parts uniformly distributed in the interval $[-0.1,0.1]$.

By using a number $n=2$ of deep and $m=4$ of hidden units per physical site, we were able to optimize the GTMS to the ground-state of a critical XXZ chain with up to $N=60$ sites, to a relative accuracy $\Delta E_{\text{rel}}=\frac{E_{_{\text{GS}}}-E_{\boldsymbol{w}}}{E_{_{\text{GS}}}}$ of $\Delta E_{\text{rel}}\approx 1\times10^{-5}$. In Fig. \ref{fig:energyevoGTMS_L60}, we show an example of a VMC convergence plot for a chain of $N=60$ sites with the translation invariant GTMS ansatz (blue data points), compared with the exact values of the ground-state energy $E_{_{\text{GS}}}$ (solid orange line) and of the first excited state energy (solid yellow line in the inset). In the inset, a comparison with the performance of a translation invariant MPS ansatz is shown (light blue data points), obtained by removing the long-range RBM couplings from the GTMS, but still optimizing the MPS using VMC. We point out that the exact value of $E_{_{\text{GS}}}$ is calculated up to an error (truncation error) of $10^{-12}$ using DMRG optimization of a MPS with open boundaries (still for the same Hamiltonian with periodic boundaries) and bond dimension of $\chi=2000$, which yields a number of variational parameters much higher than the $N_w=263$ weights parametrizing our GTMS. We also compared our GTMS ansatz with MPS of smaller bond dimension optimized with DMRG, so as to match the number of variational parameters between the two approaches for a more direct comparison. For the different MPS used we could converge to a relative accuracy on the order of $10^{-3}$. Thus the addition of the long-range couplings in the GTMS ansatz is able to improve the accuracy of at least one order of magnitude. The results are summarized in Table \ref{table:GTMS_MPS_comparison}. We point out that the GTMS results shown here, despite already yielding quite close to exact energies, may not be optimally converged, i.e. not represent a global minimum of the GTMS ansatz. As an indication in this direction, we noticed that similar precision can be achieved when only using RBM couplings, even though the GTMS states with the lowest energy are not similar to RBM states (i.e. the MPS weights are significant). We hence expect that the GTMS can yield even lower energies if the optimizer algorithm is further improved.

\begin{table}
	\begin{tabular}{l | c | c | c}
		\toprule[0.1pt]
		Ansatz  &  Optimizer  &  No. of parameters & $\Delta E_{\text{rel}}$ \\ [1.0ex] 
		
		GTMS (PB)  &  VMC  &  $263$ (total) & $1\times10^{-5}$ \\
		MPS $\chi=4$ (PB)  &  VMC  &  $27$ (total) & $1.5\times10^{-3}$ \\
		MPS $\chi=16$ (OB)  &  DMRG  &  $2\times256$ (per site) & $2.4\times10^{-3}$ \\
		\bottomrule[0.1pt]
	\end{tabular} 
	\caption{Comparison between GTMS and MPS ansatz for the critical XXZ chain of Eq. (\ref{eq:XXZhamiltonian}) with PBC and $N=60$ sites. In the first column, PB indicates that periodic boundary conditions and translation invariance are implemented in the variational parameters, while OB indicates open boundaries in the MPS ansatz, and hence no translation invariance in the parameters.}
	\label{table:GTMS_MPS_comparison}
\end{table}

\section{CONCLUDING DISCUSSION}
\label{sectio:Conclusion}
In summary, we proposed a deep ANN architecture whose the auxiliary layers can be analytically traced over and yields a class of quantum states called GTMS. The GTMS family is shown by means of a constructive mapping to include both generic MPS and RBM states, and allows to continuously interpolate between them. More specifically, GTMS networks are a family of deep Boltzmann machine networks, from which the wave function can be exactly and efficiently calculated by means of a transfer matrix method. Our findings are corroborated by numerical data showing that the GTMS network is indeed able of efficiently parametrizing random MPS, where efficiently means that the number of variational weights scales as the number of independent parameters of the MPS. Moreover we show with a numerical analysis of the second R\'enyi entropy, that GTMS initially parametrizing a MPS (therefore a state with area-law entanglement) can, upon addition of RBM weights, encode long-range correlations with volume-law entanglement.  On a general note, representation theorems \cite{Bengio_RBMrepresentability} tell us that the proposed augmentation of an existing network by additional couplings can only improve the capabilities of the network in representing quantum states. In our present construction, the onset of the volume-law scaling provides a concrete intuition for this increased representational power compared to conventional MPS. 

The potential of RBM of representing states with volume-law entanglement and encoding up to $N$-body correlations was already discussed \cite{DengVolumelaw}, and the general correspondence between RBM states and tensor network states is well known \cite{HuangPiP,GlasserMunich,Clark,ChenANN_TN}. However, an efficient mapping between MPS and RBM states has remained elusive, since it is unclear how the required number of RBM couplings scales with the bond dimension. Here, we have used the higher representational power of DBMs to efficiently and constructively embed MPS into the general framework of ANN states. By efficiently, we mean that the number of DBM couplings needed scales as the number of free parameters in the MPS tensors. In this sense, the GTMS combines key representational properties of MPS or RBM states and has stronger representational power than either of the two alone.

We explicitly demonstrated the representational powers of GTMS by learning the ground state of a critical XXZ chain with periodic boundary conditions, showing that the ground state energy is indeed correctly reproduced to a good accuracy.

The advantage of GTMS is that both short-ranged MPS correlations and long-ranged entanglement can be efficiently captured. This makes GTMS a promising ansatz for problems where the entanglement growth poses severe limitations to tensor network studies. Such cases include the study of critical systems or of time evolution in quantum many-body systems far from equilibrium, where the MPS ansatz would require us to increase the bond dimension with system size and time, respectively \cite{EisertPlenio_AreaLaw,Hastings_AreaLaw,Vedral_entanglement_rev,Schollwoeck_DMRG_Luttinger,Vidal_Kitaev_entangCritical,Laflorencie_Entang1Dcrit,Vidal_TimeEvo,White_TimeEvo,ZaletelExactMPS}. However, it is fair to say that the generalization from MPS to GTMS comes at a price: While observables can be efficiently represented directly in the space of MPS \cite{SchollwoeckMPS_DMRG,Cirac_MPS_DMRG}, for most variational wave functions including RBM states and also the proposed GTMS the understanding of the corresponding variational space is far less complete. Therefore, evaluating expectation values of physical observables and optimizing the variational parameters so far requires stochastic methods, which can pose particularly severe limitations to the accuracy of real-time evolution with VMC algorithms.

Finally, we note that various previous studies have explored VMC optimization techniques applied to the tensor network ansatz \cite{SandvikVMC_MPS,WangVMC_PEPS,FerrisVMC_MERA,SikoraVMC_PEPSprojector}. Going beyond conventional VMC in order to develop new methods for the optimization of ANN states, e.g., drawing intuition from tensor network methods such as DMRG, TEBD, and TRG, is an interesting direction of future research.

\section{ACKNOWLEDGEMENTS}
We acknowledge discussions with Emil Bergholtz, Hong-Hao Tu, Christian Mendl and Irene Lopez. LP and JCB acknowledge financial support from the German Research Foundation (DFG) through the Collaborative Research Centre SFB 1143. RK is supported by the Austrian Science Fund SFB FoQuS (FWF Project No.~F4016-N23) and the European Research Council (ERC) Synergy Grant UQUAM. The numerical calculations were performed on resources at the TU Dresden Center for Information Services and High Performance Computing (ZIH), and at the Chalmers Centre for Computational Science and Engineering (C3SE) provided by the Swedish National Infrastructure for Computing (SNIC). The DMRG optimization were done using the ITensor libraries (\href{http://itensor.org}{http://itensor.org}).

\appendix
\section{GENERAL EXACTLY CONTRACTIBLE DBM NETWORK}
\label{sectio:exactlycontractibleGTMS}
The network shown in Fig.~\ref{fig:GTMSfull} is not the most general architecture from which the wave function can be exactly calculated. As discussed in Section \ref{sectio:GTMS}, one can introduce arbitrary (from two-body to $n$-body) couplings between deep variables in the same block and in neighboring blocks by still keeping the auxiliary layers exactly traceable. In this appendix, we want to elaborate more on this structure, explaining how the presence of an hidden layer is fundamental for having enough variational parameters to parametrize generic MPS.

The network energy for a GTMS where arbitrary couplings between sets of deep variables within the same block and between neighboring blocks have been introduced, is obtained simply by adding a term
\begin{equation}
-\sum_{j=1}^{N_T}\bigg\{C_j^{\text{(o.s.)}}(\boldsymbol{d}_j)+C_j^{\text{(n.n.)}}(\boldsymbol{d}_j,\boldsymbol{d}_{j+1})\bigg\} 
\end{equation}
to the espression of $\mathcal{E}_{\text{nw}}$ in Eq.~(\ref{eq:GTMS_netwenergy}), where $C_j^{\text{(o.s.)}}(\boldsymbol{d}_j)$ denotes the sum of all possible direct couplings between the deep spins contained in block $j$:
\begin{equation}
\begin{split}
C_j^{\text{(o.s.)}}(\boldsymbol{d}_j)=\sum_{\nu_1\ne\nu_2}\tilde{\omega}_j^{\nu_1,\nu_2}d_j^{\nu_1}d_j^{\nu_2}\\
+\sum_{\nu_1\ne\nu_2\ne\nu_3}\tilde{\omega}_j^{\nu_1,\nu_2,\nu_3}d_j^{\nu_1}d_j^{\nu_2}d_j^{\nu_3}+... \,\,,
\end{split}
\end{equation}
and $C_j^{\text{(n.n.)}}(\boldsymbol{d}_j,\boldsymbol{d}_{j+1})$ denotes the sum of all possible direct couplings between the deep spins in neighboring blocks $j$ and $j+1$:
\begin{equation}
\begin{split}
C_j^{\text{(n.n.)}}(\boldsymbol{d}_j,\boldsymbol{d}_{j+1})=\sum_{\nu_1,\nu_2}\hat{\omega}_{j,j+1}^{(\nu_1)(\nu_2)}d_j^{\nu_1}d_{j+1}^{\nu_2}\\
+\sum_{\nu_1,\nu_2\ne\nu_3}\hat{\omega}_{j,j+1}^{(\nu_1)(\nu_2,\nu_3)}d_j^{\nu_1}d_{j+1}^{\nu_2}d_{j+1}^{\nu_3}\\
+\sum_{\nu_1\ne\nu_2,\nu_3}\hat{\omega}_{j,j+1}^{(\nu_1,\nu_2)(\nu_3)}d_j^{\nu_1}d_j^{\nu_2}d_{j+1}^{\nu_3}+... \,\,.
\end{split}
\end{equation}
The addition of these couplings does maintain the auxiliary layers exactly traceable. The only modification to the wave function amplitude is that the product elements $t_{j,j+1}$ of Eq.~(\ref{eq:GTMS_t_product_elements}) for block $j$ are now multiplied by the corresponding factors $\text{e}^{C_j^{\text{(o.s.)}}+C_j^{\text{(n.n.)}}}$:
\begin{equation*}
t_{j,j+1}(\boldsymbol{\sigma},\boldsymbol{d}_j,\boldsymbol{d}_{j+1})\to\text{e}^{C_j^{\text{(o.s.)}}+C_j^{\text{(n.n.)}}}t_{j,j+1}(\boldsymbol{\sigma},\boldsymbol{d}_j,\boldsymbol{d}_{j+1}) \,\,.
\end{equation*}
One may now ask the question whether, with the addition of such arbitrary links between deep variables, the number of variational parameters is sufficient to parametrize arbitrary MPS \emph{without} the addition of the hidden layer. It turns out that this is not the case. With $n$ deep spin variables per block, the number $N_{\omega}\big|_j$ of direct links, and therefore of complex weights $\tilde{\omega}_j^{\{\nu_i\}}$ and $\hat{\omega}_{j,j+1}^{(\{\nu_i\})(\{\nu_k\})}$ in the sum $C_j^{\text{(o.s.)}}+C_j^{\text{(n.n.)}}$ between blocks $j$ and $j+1$ is
\begin{equation*}
N_{\omega}\big|_j=\sum_{k=2}^{2n}\binom{2n}{k}-\sum_{k=2}^{n}\binom{n}{k}=2^{2n}-2^{n}-n-1 \,\,.
\end{equation*}
Therefore, with the addition of the bias terms for the $n$ deep spin variables, the total number of complex weights per transfer matrix in absence of hidden spin variables, would be $2^{2n}-2^{n}+n-1$, insufficient to parametrize an arbitrary MPS tensor, for which in general $2^{2n}(2s+1)$ would be required ($2s+1$ being the local Hilbert space dimension). If we require the network to be able to parametrize generic MPS, we therefore must add additional hidden spins to be traced out before the sum over the deep variables configurations is computed, in order to have enough variational parameters.

\section{OPEN BOUNDARY CONDITIONS}
\label{sectio:GTMSwithOBC}
Here we briefly mention how the sum over hidden and deep variables of a GTMS can be exactly computed without the use of PBC, thereby allowing for the parametrization of MPS with open boundary conditions and their generalization to a GTMS that interpolates between them and RBM states.

Considering the network studied in Sec. \ref{sectio:GTMS}, described by the network energy of Eq.~(\ref{eq:GTMS_netwenergy}), we set open boundary conditions on it simply by erasing (i.e. setting to $0$) the $\hat{w}$ couplings extending from tensor $j=N_T$ to $j=1$. By applying the transfer matrix method in order to calculate  the trace over hidden and deep layers of a GTMS with open boundaries we find an exact expression for the wave function, which reads as
\begin{equation}
\psi_{\boldsymbol{w}}(\boldsymbol{\sigma})=\text{e}^{\sum_ic_i\sigma_i}\bar{V}_1\bigg(\prod_{j=2}^{N_T-1}T_{j}\bigg)V_{N_T} \,\,.
\end{equation}
The transfer matrices $T_j$ in the bulk of this open boundary GTMS are the same as the ones defined in Eqs.~(\ref{eq:GTMS_t_product_elements})-(\ref{eq:deftransfer}).
The right-boundary tensor ($j=N_T$) is a $2^n\times1$ column vector with elements
\begin{equation}
\big(V_{N_T}\big)_{\boldsymbol{d}_{N_T}}=\text{e}^{\sum_{\nu}a_{N_T}^{\nu}d_{N_T}^{\nu}}\prod_{\mu=1}^m 2\cosh\big(\tilde{\varphi}_{N_T}^{\mu}(\boldsymbol{\sigma},\boldsymbol{d}_{N_T})\big) \,\,,
\end{equation}
where
\begin{equation}
\tilde{\varphi}_{N_T}^{\mu}(\boldsymbol{\sigma},\boldsymbol{d}_{N_T})=b_{N_T}^{\mu}+\sum_{i=1}^N\sigma_iw_{i,{N_T}}^{\mu}+\sum_{\nu=1}^n\tilde{w}_{N_T}^{\mu,\nu}d_{N_T}^{\nu} \,\,.
\end{equation}
The left-boundary tensor ($j=1$) is given by
\begin{equation}
\bar{V}_1=\begin{bmatrix}
1 & \cdots & 1
\end{bmatrix}T_{1} \,\,,
\end{equation}
that is a $1\times2^n$ row vector, where $T_1$ is the transfer matrix matrix with elements $t_{1,2}(\boldsymbol{\sigma},\boldsymbol{d}_1,\boldsymbol{d}_{2})$.

\section{STRING-BOND STATES FROM GTMS}
\label{sectio:SBS}
Given the ability of the GTMS to parametrize arbitrary MPS, it is easy to construct a higher-dimensional generalization of the GTMS network of Fig. \ref{fig:GTMSfull} that still allows for exact traceability of the auxiliary layers and is able to parametrize arbitrary string-bond states (SBS). As in the MPS case, the number of variational parameters required would scale with the number of parameters describing the SBS. The addition of long-range RBM couplings to such network would then result in a nonlocal dependence of the SBS matrices on the entire spin configuration of the lattice, as in the case of MPS, and would then allow the resulting state to capture entanglement beyond area law. This simple observation extends the range of applicability of our GTMS to higher-dimensional systems as well.
A SBS \cite{Schuch_Cirac_SBS} is constructed as a product of different MPS \emph{strings} which \emph{patch} a two (or higher) dimensional lattice, and reads as
\begin{equation}
	\psi_{\text{SBS}}(\boldsymbol{\sigma})=\prod_{s\,\in\,\text{strings}}\bigg(\prod_{i\in s}A_s^{[i]\sigma_i}\bigg) \,\,,
\end{equation}
where $s$ is the index labeling a given MPS string and the second product runs over the lattice sites $i$ contained in the string $s$. In systems with PBC, for strings that wrap around the lattice the trace of the above second product of matrices has to be taken, otherwise the first and last tensors of a string need to be a row- and column- vectors, as usual. In order to construct a GTMS-like DBM architecture which still allows for exact traceability of hidden and deep layers, and yields SBS after this trace is performed, we simply assign an independent one-dimensional GTMS network to each of the strings. Since the GTMS networks for each string are independent, we can still apply the transfer matrix method explained in Sec. \ref{sectio:GTMS} to trace over hidden and deep layers in each of them separately, obtaining a SBS in the special case where the long-range RBM couplings are absent. The RBM links can arbitrarily couple hidden and physical units of different strings, still maintaining the exact traceability of the hidden layers, and the quantum state resulting from them would be a \emph{generalized} SBS where the string tensors would depend nonlocally on the spin configuration of the whole lattice.

\section{EFFICIENT CALCULATION OF SECOND R\'ENYI ENTROPY}
\label{sectio:RenyiEntropyVMC}
Here we review the algorithm for the computation of the second R\'enyi entropy applicable to Monte Carlo calculations, introduced in Ref.~\cite{Hastings_RenyiEntropy}. Consider a system $S$ in a quantum state $|\psi\rangle$, and a bipartition of it into two subsystems $A$ and $B$. The R\'enyi entropy of $A$ subsystem is given by
\begin{equation}
S^{(A)}_2=-\ln\big(\text{tr}\,\rho_A^2\big) \,\,,
\end{equation}
with $\rho_A=\text{tr}_B\,|\psi\rangle\langle\psi|$ being the reduced density matrix of subsystem $A$. One can re-express $S^{(A)}_2$ in a form which is convenient for Monte Carlo calculation by considering an identical copy $S'$ of the system $S$ with the same bipartition into subsystems $A'$ and $B'$, and defining a Swap$_A$ operator acting on the tensor product of the Hilbert spaces of the two copies, which swaps the configurations in $A$ and $A'$. More concretely, let $|\boldsymbol{\sigma}\rangle$ be a state in the coordinate basis of $S$ (a spin configuration) which can be written as $|\boldsymbol{\sigma}\rangle=|\boldsymbol{\sigma}_A\boldsymbol{\sigma}_B\rangle$, where $\boldsymbol{\sigma}_A$ and $\boldsymbol{\sigma}_B$ are configurations in $A$ and $B$ respectively. Similarly, $|\boldsymbol{\sigma}'\rangle=|\boldsymbol{\sigma}'_{A'}\boldsymbol{\sigma}'_{B'}\rangle$ in $S'$. The swap operator acts on the tensor product of the two copies as $\text{Swap}_A\big(|\boldsymbol{\sigma}\rangle\otimes|\boldsymbol{\sigma}'\rangle\big)=|\boldsymbol{\sigma}'_{A'}\boldsymbol{\sigma}_{B}\rangle\otimes|\boldsymbol{\sigma}_{A}\boldsymbol{\sigma}'_{B'}\rangle$. Using this definition it is easy to show that the second R\'enyi entropy can be rewritten as
\begin{equation}
S^{(A)}_2=-\ln\big(\text{tr}\,\rho_A^2\big)=-\ln\,\langle\text{Swap}_A\rangle \,\,,
\end{equation}
where the expectation value $\langle\text{Swap}_A\rangle$ is taken over the product state $|\psi\rangle\otimes|\psi\rangle$ of the two copies. This expectation value reads
\begin{equation}
\langle\text{Swap}_A\rangle=\sum_{\{\boldsymbol{\sigma}\},\{\boldsymbol{\sigma}'\}}P(\boldsymbol{\sigma})P(\boldsymbol{\sigma}')\frac{\langle\boldsymbol{\sigma}'_{A'}\boldsymbol{\sigma}_{B}|\psi\rangle\langle\boldsymbol{\sigma}_{A}\boldsymbol{\sigma}'_{B'}|\psi\rangle}{\langle\boldsymbol{\sigma}_{A}\boldsymbol{\sigma}_{B}|\psi\rangle\langle\boldsymbol{\sigma}'_{A'}\boldsymbol{\sigma}'_{B'}|\psi\rangle} \,\,,
\label{eq:swap_expectval}
\end{equation}
where, as before, $|\boldsymbol{\sigma}\rangle=|\boldsymbol{\sigma}_{A}\boldsymbol{\sigma}_{B}\rangle$ in $S$, $|\boldsymbol{\sigma}'\rangle=|\boldsymbol{\sigma}'_{A'}\boldsymbol{\sigma}'_{B'}\rangle$ in $S'$, and $P(\boldsymbol{\sigma})=\frac{|\langle\boldsymbol{\sigma}|\psi\rangle|^2}{\langle\psi|\psi\rangle}$ is the probability density for the $\boldsymbol{\sigma}$ configuration. The double sum in Eq.~(\ref{eq:swap_expectval}) can be replaced by sum over two sets of Monte Carlo samples of $P(\boldsymbol{\sigma})$ allowing for an efficient calculation of $S^{(A)}_2$.

\section{TRANSLATION INVARIANT GTMS}
\label{sectio:translinvGTMS}
We discuss here how to implement translation invariance in a GTMS network. We start our discussion from the case of a translation invariant MPS parametrized by a GTMS network. For a translation invariant MPS, the individual tensors $A^{[i]\sigma_i}_{a_{i-1}a_i}$ at each lattice site $i$ are the independent of the site index $i$, namely, $A^{[i]\sigma_i}_{a_{i-1}a_i}=A^{\sigma_i}_{a_{i-1}a_i}$. This suggests that a GTMS network parametrizing a translation invariant MPS must as well have the weights (the black and purple links in Fig.~\ref{fig:GTMSfull}) independent of the site index, that is $w_{i,i}^{\mu}=w^{\mu}$, $\tilde{w}_i^{\mu,\nu}=\tilde{w}^{\mu,\nu}$, $\hat{w}_{i,i+1}^{\mu,\nu}=\hat{w}^{\mu,\nu}$, $c_i=c$, $b_i^{\mu}=b^{\mu}$ and $a_i^{\nu}=a^{\nu}$ (and $w_{i,j\ne i}^{\mu}=0$). Therefore, for a translation invariant MPS it is sufficient to calculate the matrices $A^{\sigma_i}_{a_{i-1}a_i}$ for the different values of $\sigma_i$ once.

On the top of the MPS weights, we can then add non-zero RBM weights $w_{i,j\ne i}^{\mu}$ in such a way that translation invariance is preserved. For this it is sufficient to set the weights $w_{i,j\ne i}^{\mu}$ as dependent only on the distance $|i-j|$ between the the physical site $i$ and the position of the tensor $j$ connected by the link. This means $w_{i,i+d}^{\mu}=w_d^{\mu}$, where one has to apply PBC by setting $i+N=i$ if $N$ is the number of physical sites of the system. However, since now the transfer matrices $T_j(\boldsymbol{\sigma})$ depend in general on the spin configuration on the whole system, one still needs to calculate all of the $T_j$ transfer matrices for each spin configuration.

\section{OPTIMIZATION OF GTMS}
\label{sectio:StochReconf}
We provide here some details on the optimization of translation invariant GTMS which we used in our VMC simulations in the main text. For the minimization of the energy expectation value we adopted the stochastic reconfiguration (SR) method, which can be understood as an imaginary time evolution projected onto the variational GTMS manifold, where at each imaginary time-step the $N_w$ variational parameters are iteratively updated. Interpreting $\boldsymbol{w}$ as a vector with $2N_w$ real components corresponding to the real and imaginary parts of the variational parameters, the SR update of the weights is calculated from 
\begin{equation}
d\boldsymbol{w}=-\gamma\,S_{\boldsymbol{w}}^{-1}\boldsymbol{F}_{\boldsymbol{w}} \,\,,
\end{equation}
where $\gamma$ is the learning rate, playing here the role of an imaginary time-step, $S_{\boldsymbol{w}}$ is the $2N_w\times 2N_w$ \emph{local metric tensor} with elements given by
\begin{equation}
\big(S_{\boldsymbol{w}}\big)_{j,k}=\langle\partial_{w_j}\hat{\psi}_{\boldsymbol{w}}|\partial_{w_k}\hat{\psi}_{\boldsymbol{w}}\rangle-\langle\partial_{w_j}\hat{\psi}_{\boldsymbol{w}}|\hat{\psi}_{\boldsymbol{w}}\rangle\langle\hat{\psi}_{\boldsymbol{w}}|\partial_{w_k}\hat{\psi}_{\boldsymbol{w}}\rangle \,\,,
\end{equation}
with $|\hat{\psi}_{\boldsymbol{w}}\rangle$ is the normalized variational GTMS state $|\psi_{\boldsymbol{w}}\rangle$, and $\boldsymbol{F}_{\boldsymbol{w}}$ is the \emph{force vector} with $2N_w$ components given by
\begin{equation}
\big(\boldsymbol{F}_{\boldsymbol{w}}\big)_j=\langle\partial_{w_j}\hat{\psi}_{\boldsymbol{w}}|H|\partial_{w_k}\hat{\psi}_{\boldsymbol{w}}\rangle-\langle\partial_{w_j}\hat{\psi}_{\boldsymbol{w}}|\hat{\psi}_{\boldsymbol{w}}\rangle\langle\hat{\psi}_{\boldsymbol{w}}|H|\hat{\psi}_{\boldsymbol{w}}\rangle \,\,.
\end{equation}
At each optimization step the above quantities are calculated with Monte Carlo sampling of the (non normalized) probability density $P_{\boldsymbol{w}}(\boldsymbol{\sigma})=|\langle\boldsymbol{\sigma}|\psi_{\boldsymbol{w}}\rangle|^2$, using 
\begin{align}
&\big(S_{\boldsymbol{w}}\big)_{j,k}=\langle O_j^*O_k\rangle_{\text{MC}}-\langle O_j^*\rangle_{\text{MC}}\langle O_k\rangle_{\text{MC}} \,\,, \\
&\big(\boldsymbol{F}_{\boldsymbol{w}}\big)_j=\langle O_j^*E_{\text{loc}}\rangle_{\text{MC}}-\langle O_j^*\rangle_{\text{MC}}\langle E_{\text{loc}}\rangle_{\text{MC}} \,\,,
\end{align}
where $\langle\cdot\rangle_{\text{MC}}$ denotes the average over the samples as Monte Carlo estimate, $E_{\text{loc}}(\boldsymbol{\sigma})=\frac{\langle\boldsymbol{\sigma}|H|\psi_{\boldsymbol{w}}\rangle}{\langle\boldsymbol{\sigma}|\psi_{\boldsymbol{w}}\rangle}$ is the local energy estimator, and the
\begin{equation}
O_j(\boldsymbol{\sigma})=\frac{\partial}{\partial w_j}\ln\big(\langle\boldsymbol{\sigma}|\psi_{\boldsymbol{w}}\rangle\big)
\end{equation}
are the local estimators for the derivatives with respect to the variational parameters. We refer the reader to Refs.~\cite{CarleoTroyer,our_article} for more practical details on the implementation. In the case of a translation invariant GTMS, the above logarithmic derivatives take the general form (apart from the derivative with respect of real and imaginary parts of $c$ which give $\sum_i\sigma_i$ and $i\sum_i\sigma_i$, respectively) 
\begin{equation}
O_j(\boldsymbol{\sigma})=\frac{e^{c\sum_i\sigma_i}}{\psi_{\boldsymbol{w}}(\boldsymbol{\sigma})}\sum_{i=1}^N\text{tr}\bigg[\Big(\prod_{\ell<i}T_{\ell}\Big)\frac{\partial T_i}{\partial w_j}\Big(\prod_{\ell>i}T_{\ell}\Big)\bigg] \,\,,
\end{equation}
where the matrix derivatives have elements
\begin{align}
&\frac{\partial\,(T_i)_{\boldsymbol{d}_i,\boldsymbol{d}_{i+1}}}{\partial\,\text{Re}(a^{\nu})}=(T_i)_{\boldsymbol{d}_i,\boldsymbol{d}_{i+1}}d_i^{\nu}\,, \\
&\frac{\partial\,(T_i)_{\boldsymbol{d}_i,\boldsymbol{d}_{i+1}}}{\partial\,\text{Re}(b^{\mu})}=(T_i)_{\boldsymbol{d}_i,\boldsymbol{d}_{i+1}}\tanh\big(\varphi_i^{\mu}(\boldsymbol{\sigma},\boldsymbol{d}_i,\boldsymbol{d}_{i+1})\big)\,, \\
&\frac{\partial\,(T_i)_{\boldsymbol{d}_i,\boldsymbol{d}_{i+1}}}{\partial\,\text{Re}(w^{\mu}_{d})}=(T_i)_{\boldsymbol{d}_i,\boldsymbol{d}_{i+1}}\sigma_{i+d}\tanh\big(\varphi_i^{\mu}(\boldsymbol{\sigma},\boldsymbol{d}_i,\boldsymbol{d}_{i+1})\big)\,, \\
&\frac{\partial\,(T_i)_{\boldsymbol{d}_i,\boldsymbol{d}_{i+1}}}{\partial\,\text{Re}(\tilde{w}^{\mu,\nu})}=(T_i)_{\boldsymbol{d}_i,\boldsymbol{d}_{i+1}}d_i^{\nu}\tanh\big(\varphi_i^{\mu}(\boldsymbol{\sigma},\boldsymbol{d}_i,\boldsymbol{d}_{i+1})\big)\,, \\
&\frac{\partial\,(T_i)_{\boldsymbol{d}_i,\boldsymbol{d}_{i+1}}}{\partial\,\text{Re}(\hat{w}^{\mu,\nu})}=(T_i)_{\boldsymbol{d}_i,\boldsymbol{d}_{i+1}}d_{i+1}^{\nu}\tanh\big(\varphi_i^{\mu}(\boldsymbol{\sigma},\boldsymbol{d}_i,\boldsymbol{d}_{i+1})\big)\,,
\end{align}
when taken with respect to the real parts of the weights, and which need just to be multiplied by $i$ if taken with respect to the imaginary part.

\bibliographystyle{apsrev}

\end{document}